\documentclass[10pt,conference]{IEEEtran}

\usepackage{cite}
\usepackage{amsmath,amssymb,amsfonts}
\usepackage{algorithmic}
\usepackage{graphicx}
\usepackage{textcomp}
\usepackage[hyphens]{url}
\usepackage[table]{xcolor}
\def\BibTeX{{\rm B\kern-.05em{\sc i\kern-.025em b}\kern-.08em
    T\kern-.1667em\lower.7ex\hbox{E}\kern-.125emX}}

\usepackage{hyperref}
\hypersetup{hidelinks,colorlinks=true,linkcolor=red,citecolor=purple}
\usepackage[numbers,sort,compress]{natbib}

\usepackage{soul}

\usepackage{epsfig}
\usepackage{times}
\usepackage[ruled]{algorithm2e}
\usepackage{bbding}
\usepackage{pifont}
\usepackage{wasysym}
\usepackage{latexsym}

\usepackage{nonfloat}
\usepackage{multirow}

\renewenvironment{thebibliography}[1]{%
\begin{oldthebibliography}{#1}%
  \setlength{\parskip}{0ex}%
  \setlength{\itemsep}{0ex}%
}%
{%
\end{oldthebibliography}%
}

\usepackage{listings}
  \usepackage{courier}
\usepackage[textsize=tiny,textwidth=0.6in]{todonotes}
\newcommand{\allnotes}[1]{}
\renewcommand{\allnotes}[1]{#1} 

\renewcommand{\em}{\it}

\newcommand{\x}{$\times$}

\newcommand{\ignore}[1]{}

\def\cfigure[#1,#2,#3]{
\begin{figure}
\vspace*{0mm}
\begin{center}

\includegraphics[width=3in]{#1} 
 
\vspace*{-3mm}\caption[]{#2
} \label{#3}
 
\vspace*{-5mm}
\end{center}
\end{figure}}

\def\cfigurefour[#1,#2,#3]{
\begin{figure}
\vspace*{0mm}
\begin{center}

\includegraphics[width=4in]{#1} 
 
\vspace*{-3mm}\caption[]{#2
} \label{#3}
 
\vspace*{-5mm}
\end{center}
\end{figure}}

\def\cfiguretemp[#1,#2,#3]{
\begin{figure}
\vspace*{0mm}
\begin{center}

\includegraphics[width=3.5in]{#1} 
 
\vspace*{-3mm}\caption[]{#2
} \label{#3}
 
\vspace*{-5mm}
\end{center}
\vspace*{-2mm}
\end{figure}}

\def\wfigure[#1,#2,#3]{
\begin{figure*}
\vspace*{0mm}
\begin{center}
 \includegraphics[width=\textwidth]{#1} 
 \vspace*{-3mm}\caption[]{#2
} \label{#3}
 
\end{center}
\end{figure*}}

\def\threefigure[#1,#2,#3,#4,#5]{
\begin{figure*}
\vspace*{0mm}
\begin{center}

\begin{tabular}{ccc}
\includegraphics[width=2in]{#1} & \includegraphics[width=2in]{#2} &  \includegraphics[width=2in]{#3} \\
(a) & (b) & (c) \\
\end{tabular}

\vspace*{-3mm}\caption[]{#4
} \label{#5}

\vspace*{-5mm}
\end{center}
\vspace*{-2mm}
\end{figure*}}

\def\dcfigure[#1,#2,#3,#4,#5,#6]{
{
\begin{figure*}
\begin{center}
\begin{minipage}[c]{\columnwidth}{
\includegraphics[width=\columnwidth]{#1} 
\vspace*{0mm}\caption[]{#2} \label{#3} \
}\end{minipage}\hspace*{\columnsep}\
\begin{minipage}[c]{\columnwidth}{
\includegraphics[width=\columnwidth]{#4} 
\vspace*{0mm}\caption[]{#5}\label{#6} \
}\end{minipage}
\end{center}
\end{figure*}
}
}

\def\tableByTable[#1,#2,#3,#4,#5,#6]{
{
\begin{table*}
\begin{center}
\begin{minipage}[c]{3in}{
\centering
{#1}
\vspace*{0mm}\tabcaption[]{#2}\label{#3} \
}\end{minipage}\hspace*{\columnsep}\
\begin{minipage}[c]{3in}{
\centering
{#4}
\vspace*{0mm}\tabcaption[]{#5}\label{#6} \
}\end{minipage}
\end{center}
\end{table*}
}
}

\def\figureByTable[#1,#2,#3,#4,#5,#6]{
{
\begin{figure*}
\begin{center}
\begin{minipage}[c]{3in}{
\centering
\includegraphics[width=\textwidth]{#1}
\vspace*{0mm}\figcaption[]{#2} \label{#3} \
}\end{minipage}\hspace*{\columnsep}\
\begin{minipage}[c]{3.3in}{
\centering
{#4}
\vspace*{0mm}\tabcaption[]{#5}\label{#6} \
}\end{minipage}
\end{center}
\end{figure*}
}
}

\def\tableByFigure[#1,#2,#3,#4,#5,#6]{
{
\begin{figure*}
\begin{center}
\begin{minipage}[c]{4.3in}{
\centering
{#1}
\vspace*{0mm}\tabcaption[]{#2} \label{#3} \
}\end{minipage}\hspace*{\columnsep}\
\begin{minipage}[c]{2.2in}{
\centering
\includegraphics[width=\textwidth]{#4}
\vspace*{-0.35in}\caption[]{#5}\label{#6} \
}\end{minipage}
\end{center}
\end{figure*}
}
}

\def\doublecfigure[#1,#2,#3,#4]{
{
\begin{figure}
\begin{center}
\begin{minipage}[c]{1.5in}{
\begin{center}
\includegraphics[width=1.5in]{#1}
\end{center}
}\end{minipage}\hspace*{1em}\
\begin{minipage}[c]{1.5in}{
\begin{center}
\includegraphics[width=1.5in]{#2}
\end{center}
}\end{minipage}
\vspace*{0mm}\caption[]{#3} \label{#4} \
\end{center}
\end{figure}
}
}

\def\qcfigure[#1,#2,#3,#4,#5,#6]{
{
\begin{figure*}
\vspace*{0.2in}\
\begin{center}
\begin{minipage}[c]{3in}{
\includegraphics[width=3in]{#1} 
\vspace*{-3mm}
}
\end{minipage}\hspace*{0.5in}\
\begin{minipage}[c]{3in}{
\includegraphics[width=3in]{#2} 
\vspace*{-3mm}
}\end{minipage}

\begin{minipage}[c]{3in}{
\includegraphics[width=3in]{#3} 
\vspace*{-3mm}
}
\end{minipage}\hspace*{0.5in}\
\begin{minipage}[c]{3in}{
\includegraphics[width=3in]{#4} 
\vspace*{-3mm}
}\end{minipage}
\end{center}
\caption[]{#5}\label{#6}
\end{figure*}
}
}

\def\twfigure[#1,#2,#3,#4,#5]{
{
\begin{figure*}
\vspace*{0.2in}\
\begin{center}
\begin{minipage}[c]{6.5in}{
\includegraphics[width=6.5in]{#1} 
\vspace*{-3mm}
}
\end{minipage}

\begin{minipage}[c]{6.5in}{
\includegraphics[width=6.5in]{#2} 
\vspace*{-3mm}
}\end{minipage}

\begin{minipage}[c]{6.5in}{
\includegraphics[width=6.5in]{#3} 
\vspace*{-3mm}
}
\end{minipage}
\end{center}
\caption[]{#4}\label{#5}
\end{figure*}
}
}

\def\dwfigure[#1,#2,#3,#4]{
{
\begin{figure*}
\vspace*{0.2in}\
\begin{center}
\begin{minipage}[c]{6.5in}{
\includegraphics[width=6.5in]{#1} 
\vspace*{-3mm}
}
\end{minipage}

\begin{minipage}[c]{6.5in}{
\includegraphics[width=6.5in]{#2} 
\vspace*{-3mm}
}\end{minipage}

\end{center}
\caption[]{#3}\label{#4}
\end{figure*}
}
}

\def\dssfigure[#1,#2,#3,#4,#5,#6]{
{
\begin{figure*}
\vspace*{0.2in}\
\begin{center}
\begin{minipage}[c]{4in}{
\includegraphics[width=4in]{#1}
\vspace*{-3mm}\caption[]{#2} \label{#3} \
}\end{minipage}\hspace*{0.5in}\
\begin{minipage}[c]{2in}{
\includegraphics[width=2in]{#4}
\vspace*{-3mm}\caption[]{#5}\label{#6} \
}\end{minipage}
\end{center}
\vspace*{-0.4in}\
\end{figure*}
}
}

\def\dsfigure[#1,#2,#3,#4,#5,#6]{
{
\begin{figure*}
\vspace*{0.2in}\
\begin{center}
\begin{minipage}[c]{3in}{
\includegraphics[width=3in]{#1}
\vspace*{-3mm}\caption[]{#2} \label{#3} \
}\end{minipage}\hspace*{0.5in}\
\begin{minipage}[c]{3in}{
\hspace*{0.5in}\
\includegraphics[height=3in]{#4}
\vspace*{-3mm}\caption[]{#5}\label{#6} \
}\end{minipage}
\end{center}
\vspace*{-0.4in}\
\end{figure*}
}
}

\def\dsyfigure[#1,#2,#3,#4,#5,#6]{
{
\begin{figure*}
\vspace*{0.2in}\
\begin{center}
\begin{minipage}[c]{2.5in}{
\includegraphics[height=2.5in]{#1}
\vspace*{-3mm}\caption[]{#2} \label{#3} \
}\end{minipage}\hspace*{0.5in}\
\begin{minipage}[c]{2.5in}{
\includegraphics[height=2.5in]{#4}
\vspace*{-3mm}\caption[]{#5}\label{#6} \
}\end{minipage}
\end{center}
\vspace*{-0.4in}\
\end{figure*}
}
}

\def\dyfigure[#1,#2,#3,#4,#5,#6]{
{
\begin{figure*}
\vspace*{0.2in}\
\begin{center}
\begin{minipage}[c]{3in}{
\includegraphics[height=3in]{#1} 
\vspace*{-3mm}\caption[]{#2} \label{#3} \
}\end{minipage}\hspace*{0.5in}\
\begin{minipage}[c]{3in}{
\includegraphics[height=3in]{#4} 
\vspace*{-3mm}\caption[]{#5}\label{#6} \
}\end{minipage}
\end{center}
\vspace*{-0.4in}\
\end{figure*}
}
}

\def\dyoldfigure[#1,#2,#3,#4,#5,#6]{
{
\begin{figure*}
\vspace*{0.2in}\
\begin{center}
\begin{minipage}[c]{3in}{
\epsfysize=2.0in\
\hspace{0.5in}\
\epsfbox{#1}
\vspace*{-3mm}\caption[]{#2} \label{#3} \
}\end{minipage}\hspace*{0.25in}\
\begin{minipage}[c]{3in}{
\epsfysize=2.0in\
\hspace{0.5in}\
\epsfbox{#4}
\vspace*{-3mm}\caption[]{#5}\label{#6} \
}\end{minipage}
\end{center}
\vspace*{-0.4in}\
\end{figure*}
}
}

\def\cfiguredouble[#1,#2,#3,#4]{
\begin{figure}
\vspace*{0.2in}\
\begin{center}
\begin{minipage}[c]{1.5in}{
\epsfxsize=1.5in\
\epsfbox{#1}
}\end{minipage}\hspace*{0.1in}\
\begin{minipage}[c]{1.5in}{
\epsfxsize=1.5in\
\vspace{0.1in}\epsfbox{#2}
}\end{minipage}\vspace*{-0.10in} \caption[]{#3}\label{#4}
\end{center}
\vspace*{-0.4in}\
\end{figure}
}

\def\wpfigure[#1,#2,#3,#4]{
\begin{figure*}
\vspace*{4mm}
\begin{center}

\includegraphics[width=#4]{#1} 

\vspace*{-3mm}\caption[]{#2
} \label{#3}

\vspace*{-5mm}
\end{center}
\end{figure*}}

\def\wprfigure[#1,#2,#3,#4,#5]{
\begin{figure*}
\vspace*{4mm}
\begin{center}

\includegraphics[width=#4, angle=#5]{#1} 

\vspace*{-3mm}\caption[]{#2
} \label{#3}

\vspace*{-5mm}
\end{center}
\end{figure*}}

\def\DoubleFigureWSlide[#1,#2,#3,#4,#5,#6,#7,#8,#9]{
\begin{figure*}
\vspace*{#9}
\begin{center}
\begin{minipage}{#4}
\includegraphics[width=#4]{#1}
\vspace*{-3mm}\caption{#2
}\label{#3}
\end{minipage}
\hspace{2em}
\begin{minipage}{#8}
\includegraphics[width=#8]{#5}
\vspace*{-3mm}\caption{#6
}\label{#7}
\end{minipage}
\vspace*{-5mm}
\end{center}
\end{figure*}
}

\def\DoubleFigureW[#1,#2,#3,#4,#5,#6,#7,#8]{
\begin{figure*}
\vspace*{0in}
\begin{center}
\begin{minipage}{#4}
\includegraphics[width=#4]{#1}
\vspace*{-3mm}\caption{#2
}\label{#3}
\end{minipage}
\hspace{2em}
\begin{minipage}{#8}
\includegraphics[width=#8]{#5}
\vspace*{-3mm}\caption{#6
}\label{#7}
\end{minipage}
\vspace*{-5mm}
\end{center}
\end{figure*}
}

\def\DoubleFigureWHack[#1,#2,#3,#4,#5,#6,#7,#8]{
\begin{figure*}
\vspace*{0in}
\begin{center}
\begin{minipage}{3in}
\includegraphics[width=#4]{#1}
\vspace*{-3mm}\caption{#2
}\label{#3}
\end{minipage}
\hspace{2em}
\begin{minipage}{3in}
\includegraphics[width=#8]{#5}
\vspace*{-3mm}\caption{#6
}\label{#7}
\end{minipage}
\vspace*{-5mm}
\end{center}
\end{figure*}
}

\def\ddcfigure[#1,#2,#3,#4]{
\begin{figure*}
\vspace*{0.2in}\
\begin{center}
\begin{minipage}[c]{\columnwidth}{
\includegraphics[width=\columnwidth]{#1} 
}\end{minipage}\hspace{0.5in}\
\begin{minipage}[c]{\columnwidth}{
\includegraphics[width=\columnwidth]{#2} 
}\end{minipage} \caption[]{#3}\label{#4}
\end{center}
\end{figure*}
}

\def\ddcfigureSlide[#1,#2,#3,#4,#5]{
\begin{figure*}
\vspace*{#5}\
\begin{center}
\begin{minipage}[c]{3in}{
\includegraphics[height=3in]{#1} 
}\end{minipage}\hspace{0.5in}\
\begin{minipage}[c]{3in}{
\includegraphics[height=3in]{#2} 
}\end{minipage}\vspace*{-0.10in} \caption[]{#3}\label{#4}
\end{center}
\vspace*{-0.4in}\
\end{figure*}
}

\def\cxfigure[#1,#2,#3]{
\begin{figure}
\vspace*{4mm}
\begin{center}
 
\epsfxsize=2.5in\
\epsfbox{#1}\
 
\vspace*{-0.10in}\caption[]{#2
} \label{#3}
 
\vspace*{-5mm}
\end{center}
\vspace*{-2mm}
\end{figure}}

\newcommand{\figWidthSix}{1.72in}

\newcommand{\beforecaption}{\vspace{-.15cm}\begin{spacing}{0.85}}
\newcommand{\aftercaption}{\vspace{-.45cm}\end{spacing}}
\newcommand{\mycaption}[3]{\caption{\label{#1}{\bf #2} \em\small #3}}

\newcommand{\eg}{\textit{e.g.}}
\newcommand{\ie}{\textit{i.e.}}

\newcommand{\KB}{\,KB}
\newcommand{\MB}{\,MB}
\newcommand{\GB}{\,GB}

\newcommand{\Gbps}{\,Gbps}

\newcommand{\mus}{\mbox{$\mu s$}}
\newcommand{\ms}{\mbox{$ms$}}

\newcommand{\boldpara}[1]{\noindent{\textbf{#1}}}
\newcommand{\bolditpara}[1]{\noindent{\underline{\textbf{#1}}}}

\newcommand{\sysname}{SuperNIC}
\newcommand{\snic}{sNIC}
\newcommand{\nt}{NT}

\usepackage{tikz}

\newif\ifremark
\long\def\remark#1{
\ifremark%
        \begingroup%
        \dimen0=\columnwidth
        \advance\dimen0 by -1in%
        \setbox0=\hbox{\parbox[b]{\dimen0}{\protect\em #1}}
        \dimen1=\ht0\advance\dimen1 by 2pt%
        \dimen2=\dp0\advance\dimen2 by 2pt%
        \vskip 0.25pt%
        \hbox to \columnwidth{%
                \vrule height\dimen1 width 3pt depth\dimen2%
                \hss\copy0\hss%
                \vrule height\dimen1 width 3pt depth\dimen2%
        }%
        \endgroup%
\fi}

\newcommand{\fixme}[1]{{\color{red}\textbf{\fbox{FIXME} #1}}}

\newcommand{\TODO}[1]{{\color{red}\textbf{\fbox{TODO} #1}}}

\newcommand{\will}[1]{{\color{olive}\textbf{\fbox{Will} #1}}}

\newcommand{\myitem}[1]{\item \textbf{#1}}

\newcommand{\revise}[1]{{#1}}
\newcommand{\novelty}[1]{{#1}}

\title{SuperNIC: A Hardware-Based, Programmable, and Multi-Tenant SmartNIC}

\author{
Yizhou Shan\textsuperscript{\textdagger},
Will Lin\textsuperscript{\textdagger},
Ryan Kosta\textsuperscript{\textdagger},
Arvind Krishnamurthy\textsuperscript{*},
Yiying Zhang\textsuperscript{\textdagger} \\
\it{
\textsuperscript{\textdagger}University of California San Diego, \textsuperscript{*}University of Washington\vspace{-0.2in}}}

\begin{document}
\maketitle
\thispagestyle{plain}
\pagestyle{plain}


\begin{abstract}

With CPU scaling slowing down in today's data centers, more functionalities are being offloaded from the CPU to auxiliary devices. One such device is the SmartNIC, which is being increasingly adopted in data centers. In today's cloud environment, VMs on the same server can each have their own network computation (or {\em network tasks}) or workflows of network tasks to offload to a SmartNIC. These network tasks can be dynamically added/removed as VMs come and go and can be shared across VMs. Such dynamism demands that a SmartNIC not only schedules and processes packets but also manages and executes offloaded network tasks for different users. Although software solutions like an OS exist for managing software-based network tasks, such software-based SmartNICs cannot keep up with the quickly increasing data-center network speed.

This paper proposes a new SmartNIC platform called \textit{SuperNIC} that allows multiple tenants to efficiently and safely offload FPGA-based network computation DAGs. For efficiency and scalability, our core idea is to group network tasks into chains that are connected and scheduled as one unit. We further propose techniques to automatically scale network task chains with different types of parallelism. Moreover, we propose a fair share mechanism that considers both fair space sharing and fair time sharing of different types of hardware resources. Our FPGA prototype of SuperNIC achieves high bandwidth, low latency performance whilst efficiently utilizing and fairly sharing resources.

\end{abstract}

\section{Introduction}
\label{sec:intro}

Data-center networking is seeing three trends recently.
First, with the slowdown of Moore's Law and Denard's Scaling, more network functionalities are offloaded from the CPU to network devices like RDMA NICs. In a cloud or virtualized data-center environment, that means many tenants will be sharing the same network device.
Second, more network devices such as SmartNICs~\cite{SmartNIC-nsdi18,lambdanic,ipipe-sigcomm19} and programmable switches~\cite{distcache-fast19,netcache-nsdi18,racksched-osdi20,RMT-SIGCOMM13} are offering programmability that allows users to offload customized network functions.
Third, network speed in the data center is increasing fast. 
Today, 40\Gbps\ and 100\Gbps\ are the norm, with 200\Gbps~\cite{Mellanox-ConnectX6} available and 400\Gbps~\cite{Mellanox-ConnectX7} on the horizon. 

As a result, we anticipate the need for a SmartNIC that offers (1) multi-tenancy support, (2) programmability, and (3) hardware acceleration.
Both (1) and (2) are essential to providing the flexibility, resource-efficiency, and safety of user network computation offloading, while (3) is essential to providing the network-line-rate performance of such offloading.
Unfortunately, no existing SmartNIC solutions offer these three features together.

In this paper, we propose \textit{SuperNIC} (or {\em \snic} for short), a hardware-based, programmable, and multi-tenant SmartNIC.
\snic\ consists of an ASIC for fixed systems logic that receives, schedules, and sends packets, an FPGA for executing user-offloaded network computation, and software cores for executing the control plane.
We support three types of network computation offloading, and we collectively call them {\em network task}s, or {\em \nt}s. 
The first type is traditional network stack capabilities running at server CPU, such as a transport layer. The second is network functions commonly seen in today's data-center network management, such as firewalls and IPSec. The third is application-specific packet processing such as key-value store operations~\cite{flexnic-asplos16,pegasus-osdi20}, real-time analytics~\cite{flexnic-asplos16}, and serverless/microservice functions~\cite{lambdanic,E3}.
In a sense, each \nt{} can be thought of as a network-oriented accelerator that is offloaded to \snic, and different tenants can dynamically choose what \nt{}s to offload for their workloads.
In addition to deploying single \nt{}s to \snic, users can deploy a DAG of \nt{}s (\ie, a directed task flow). We expect DAGs of \nt{}s to be more common as they allow users to develop their network computation in a microservice manner or to easily put together a set of existing, third-party \nt{}s~\cite{nfchain-ANCS18}.

\snic\ enables more users to deploy more types of \nt{}s, in a dynamic and more complex way. This presents an interesting and challenging new research question: \textit{\textbf{How to schedule and deploy \nt{}s?}}
Traditional network devices focus on scheduling packets or flows, while operating systems schedule software execution units. It's unclear how to best schedule and manage hardware-based network tasks. 

To answer this question, we first solve a connectivity problem: \textit{\textbf{how to connect \nt{}s to the packet scheduler}} that receives packets from receiving ports and schedules packets' execution on \snic. Prior works~\cite{panic-osdi20} connect all \nt{}s to a crossbar, which is then connected to a packet scheduler.
As the number of \nt{}s increase, this solution would require a complex crossbar that consumes a huge FPGA area and/or increase switching latency.
To solve this scalability problem, our idea is to group \nt{}s that are likely to be executed in a sequence into a {\em chain} and only use one port on the crossbar to connect to an entire chain.
We increase the flexibility of this chaining design by supporting the {\em skipping} of \nt{}s in a chain for packets that do not access the entire chain.

Next, we answer the question of \textit{\textbf{how to improve the latency and throughput of \nt\ DAG execution}} by introducing two types of parallelism in addition to packet pipelining. The first type explores the parallelism within an \nt\ DAG by executing multiple \nt\ chains in parallel so as to shorten the total execution time of an \nt\ DAG. The second type increases the overall packet execution throughput by creating multiple parallel instances of an \nt\ DAG or a subset of it. We automatically determine the type and amount of parallelism for an \nt\ DAG based on request load, \snic\ resource availability, and proper share of the resource a user gets.

The third key new problem we solve is \textit{\textbf{how to fairly and efficiently share FPGA and other NIC resources}} across multiple tenants. 
We support multiple types of resource sharing, including the {\em space sharing} of FPGA chip, {\em bandwidth sharing} of an NT chain or a part of it, and {\em time sharing} an FPGA area by context switching between multiple NT chains.
\snic\ needs to ensure fairness across tenants when performing all these types of sharing.
Traditional fairness solutions only consider space- or time-sharing. We propose an algorithm that jointly considers fair space and time sharing in an adaptive and fine-grained manner. 
We also propose techniques to avoid or hide the overhead of time sharing FPGA resources across tenants.

We prototype \snic\ with FPGA using a 100\Gbps\ HiTech Global HTG-9200 boards~\cite{htg9200}.
We build six \nt{}s in three types to run on \snic:
a reliable transport, traditional network functions like firewall and encryption, and application-specific tasks such as key-value data replication and caching.
We evaluate \snic\ with micro- and macro-benchmarks and compare \snic\ with PANIC~\cite{panic-osdi20}, a recent multi-tenant SmartNIC that supports ASIC-based and CPU-based offloads.
Our results show that sNIC is able to deliver 100\Gbps\ throughput while adding only 196 ns scheduling overhead. 
Our real \nt-DAG experiments reveal that our \nt-chain-based scheduling system can largely reduce the crossbar size while reducing \nt-DAG latency by up to 40\% compared to PANIC.
Furthermore, our \nt-sharing mechanism improves performance per FPGA area by up to 2.81x,
and our fairness algorithm achieves better aggregated utilization while guaranteeing fairness. 

\section{Motivation and Related Works}
\label{sec:motivation}


\subsection{Network Function Offloading in Data Centers}

While CPU's frequency scaling is slowing down, network speed is increasing much faster.
Today, most data centers are running at 40\Gbps\ or 100\Gbps~\cite{Aquila,hpcc-sigcomm19}.
Soon, 200\Gbps~\cite{Mellanox-ConnectX6} and 400\Gbps~\cite{Mellanox-ConnectX7} networks will arrive.
As a result, the CPU consumption of software network stacks becomes increasingly prohibitive.
Network stacks tend to consume 30-40\% of CPU cycles~\cite{tonic-nsdi20}.
As such, more network functionalities are being offloaded from the CPU to various networking devices.
For example, RDMA NICs execute a transport layer in hardware and  allow the full bypass of the CPU. 
Another example is Aquila, a recent network system developed by Google that aims for ultra-low-latency communication~\cite{Aquila}.
Aquila consists of a set of customized network devices called TiNs, each of which provides NIC-like functionality~\cite{1RMA-sigcomm20} and switching functionality. 

\revise{
Apart from the above processor trend, application needs for accelerated packet processing in a cloud environment are another driver for more powerful SmartNICs that can support different function offloads for many tenants.
The first type of needs is traditional network problems like packet scheduling~\cite{Sivaraman-SIGCOMM16}, congestion control~\cite{Sharma-NSDI17,Sharma-NSDI18}, and load balancing~\cite{Miao-SIGCOMM17}.
The second broad type is applications-specific computation such as accelerating consensus~\cite{Li-OSDI16}, storage~\cite{netcache-sosp17,netchain-nsdi18}, databases~\cite{Zhu-VLDB19, eris-sosp17,Lerner-CIDR19}, and machine learning~\cite{sapio-nsdi21}.
In a cloud environment, a physical machine can host hundreds or even thousands of (lightweight) VMs or containers, each of which could have its own network offloading need. 
As the cloud keeps adopting more lightweight virtualization environments, we expect this need to grow even more in the future, justifying the need for a device like \snic.
}

\subsection{Existing SmartNIC Solutions}
\label{sec:smartnic}

SmartNICs are NICs with programmability.
Depending on the hardware providing the programmability, SmartNICs can be categorized into three types.

The first type is SoC-based SmartNICs that run a Linux-like operating system to host user software programs, usually on an ARM processor~\cite{bluefield,leapio-asplos}.
Software is flexible but cannot sustain the high processing speed needs with today's 100\Gbps\ or higher line rate~\cite{bfvpn-iciws}. 

The second type is ASIC-based SmartNICs, which include specialized network-function accelerators such as AES, compression, regular expression matching, and flow steering/filtering.
Although ASICs often offer excellent performance, they only offer fixed sets of functionalities and cannot meet the needs of users who desire to offload customized functionalities.
Moreover, ASIC makes it hard and expensive to iterate over versions and updates of deployed network functionalities.
Because of the ASIC limitation, many recent SmartNICs combine general-purpose processors with ASIC accelerators~\cite{Agilio-SmartNIC,bluefield,url:liquidio}. For example, NVidia BlueField SmartNICs~\cite{bluefield} use general-purpose cores and several fixed-logic network function accelerators together with an RDMA NIC to support network processing offloading. Although when only using the fixed-logic accelerators, BlueField can achieve high throughput, when software offloading is added, the performance drops dramatically~\cite{osti-bf2-perf}. 

The third type is FPGA-based SmartNICs. Unlike software-based or ASIC-based SmartNICs, FPGA-based ones support full programmability at the hardware speed. Because of this benefit and with FPGA development tool chains becoming mature, FPGA and FPGA-based SmartNICs have been deployed at scale inside Microsoft~\cite{SmartNIC-nsdi18} and Alibaba~\cite{Alibaba-Xilinx} and offered as a cloud service in public clouds like AWS~\cite{AWS-F1}, Alibaba Cloud~\cite{Alibaba-FPGA}, Tencent Cloud~\cite{Tencent-FPGA}.
For the same reason and following cloud trends, we also adopt FPGA as the media for executing \nt{}s in \snic.

As more data-center workloads and cloud users use FPGA-based SmartNICs, there will be the need to share FPGA-based SmartNICs across multiple tenants. 
Unfortunately, no existing works provide multi-tenancy support for FPGA-based SmartNICs. 
This paper fills this hole by proposing \textit{\textbf{the first multi-tenant FPGA-based SmartNIC, \snic}}. 


Among all prior SmartNIC solutions, PANIC~\cite{panic-osdi20} is the most relevant to \snic. 
PANIC is a SmartNIC platform that schedules and executes chains of network functionalities for multiple tenants.
\revise{
There are four main differences between PANIC and \snic.
First, PANIC focuses on packet scheduling, while \snic\ focuses on \nt\ scheduling in addition to packet scheduling.
Second, PANIC's design is for fixed-logic network function accelerators and CPU-based compute units. In contrast, \snic\ is designed for FPGA-based SmartNICs and solves unique challenges related to FPGA space sharing and reconfiguration. Although PANIC uses FPGA for prototyping, they do not address such FPGA issues.
Third, PANIC connects all network function units directly to a crossbar, thereby incurring space and/or performance overhead and scalability limitations.
\snic\ uses the novel \nt-chain mechanism to reduce both the burden on the crossbar and the overhead of the scheduler (\S\ref{sec:nt}).
Finally, unlike \snic, PANIC only has primitive fairness support (\eg, Weighted Fair Queuing), not handling fair spatial and temporal allocation of different hardware resources.
}

\subsection{Multi-Tenancy Support for Generic FPGA Sharing}
There are several solutions that provide virtualized, isolated generic FPGA environments that can be used by multiple tenants for computing acceleration as explained below, but none target network acceleration (FPGA-based SmartNICs).  

The first sharing mechanism is time multiplexing, where an entire FPGA chip is dedicated to one tenant for a time period before it is reconfigured to serve the next tenant. Today's cloud FPGA services like AWS F1~\cite{AWS-F1} take this approach.
The main issue with this mechanism is that an FPGA chip can be largely idle when a tenant only uses a small part of it.

The second type is space sharing, where different tenants' applications run on different parts of an FPGA chip.
Initial efforts~\cite{chen-cf14,byma-fccm14,fahmy-cloudcom15,weerasinghe-uic-atc-scalcom15,rc3e-cloud16} for FPGA space sharing partition the physical FPGA into fixed sized \emph{slots} each of which is assigned exclusively to an application.
For example, Coyote~\cite{coyote-osdi20}
virtualizes FPGA by dividing it into a number of fixed-size slots and scheduling user tasks onto these slots. 
%
Another approach is exemplified by AmorphOS~\cite{amorphos-osdi18}, which packs FPGA applications that are then scheduled onto dynamically sized slots.
%
The most recent work, ViTAL~\cite{vital-asplos20}, compiles and decomposes an FPGA application into a set of fixed-size chunks, each of which can freely run on any fixed-size slots in an FPGA.
%

Although these prior works proposed various solutions to time- and space-share an FPGA, they are not targeting network usages and are largely orthogonal to \snic. 
\snic\ is a multi-tenant SmartNIC that customizes the FPGA for executing network task DAGs, by allowing an \nt\ DAG to be broken into subsets that run in different FPGA regions and by allowing different users to share a subset of \nt{}s. 
In addition to space-sharing and time-sharing with context switching, \snic\ also allows multiple tenants to safely share the same \nt's bandwidth.
We further propose different types of \nt\ parallelism and autoscaling techniques.




\subsection{Fair Sharing of Network Devices}
\label{sec:related-fairness}

As more customized network functions are offloaded to network devices, the requirements on multi-tenancy mechanisms also become higher~\cite{fairnic-sigcomm20,wang-hotcloud20}. To provide performant multi-tenancy, a key challenge that needs to be solved is performance isolation.

Performance isolation for network devices is typically provided by a packet scheduling policy that ensures each tenant gets their \emph{fair share}.
A host of solutions have been proposed to fairly share the link bandwidth of network devices~\cite{WFQ-fairness,GPS-fairness,DRR-fairness,SFQ-fairness}.
They treat a network device as a single type of resource that is time-shared by different flows.
However, today's SmartNICs have many different types of resources, such as accelerators, general-purpose cores, and on-board memory that can be shared by multiple tenants. 

DRF~\cite{DRF} is a seminal multi-resource space-sharing solution that guarantees that the ``dominant'' resources of different users get their fair space shares. 
As the input to DRF, each user specifies a vector whose element represents her demand for a particular type of resource. DRF finds the dominant type of resource and the dominant share for each user.
For example, consider a user $x$ who requires 1 CPU core and 4\GB\ of memory for a unit task. When scheduled on a server with 9 CPU cores and 18\GB\ of memory, the user's dominant resource is memory (2/9 of the total memory as opposed to 1/9 of CPU cores). DRF provides an allocation that equalizes the dominant shares of different users while maximizing resource utilization. Suppose another user $y$ demands 3 cores and 1\GB\ of memory per unit task. DRF would solve the equations of $x+3y\leq 9$, $4x+y \leq 18$, and $2x/9 = y/3$ (result being $x=3$, $y=2$). 
DRF targets a server setting and does not consider time-sharing by statically assigning each user the exclusive ownership of a part of a resource over the entire duration. It prevents a user from using any statically assigned resource to another user, even if the latter is not using all of it.

DRFQ~\cite{DRFQ} is a multi-resource fairness algorithm for time-shared resources that allows each user to get the fair time shares of her dominant resource.
\revise {
DRFQ adapts DRF with the additional notion of \emph{packet processing time}, representing the processing time of a particular packet at a resource. It aims to allocate a fair dominant share of packet processing time across users. This fair time-sharing mitigates the above problem of DRF. 
However, DRFQ is insufficient for processing hardware like FPGA that can be configured into different units that execute at the same time. DRFQ would treat FPGA as one unit and time share it. 
}

Different from these prior works, \snic\ considers both fair space and fair time sharing of its hardware resources (\S\ref{sec:policy}), \revise{which is what is needed for network processing hardware like FPGA}. 
Another difference is that \snic\ needs to consider each \nt\ DAG as a distinct type of resource, because \snic\ supports the time sharing of an \nt\ DAG.
Moreover, we propose a new approach to measuring user load requirements.

\subsection{Summary}
In short, this work is motivated by the need in data centers for multiple tenants to offload an increasing amount of network functionalities to a SmartNIC and for the offloads to be fully programmable, hardware-based, and properly isolated. 
As no such solution exists prior to this work, we fill up an important gap in the design space of SmartNICs by building \snic, a multi-tenant FPGA-based SmartNIC that achieves high and fairly-isolated performance for network task offloading.

\section{Usage Model and Design Overview}
\label{sec:overview}

Before delving into the detailed design of \snic, this section first gives an overview of \snic, how to use it, its high-level architecture, and path taken by a packet through \snic.

\subsection{Using SuperNIC}
\label{sec:interface}

To use \snic, users first write and deploy \nt{}s as FPGA netlists; they can also use provider-supplied or third-party FPGA netlists.
Optionally, users can specify what \nt{}(s) they are willing to share with other users, with \snic's guaranteed performance and memory isolation. 
These \nt{}s are usually supplied by the cloud provider or a trusted third party.
We expect users who share \nt{}s to not trust each other but trust the supplier of the shared \nt{}s. 
We expect most sharing cases to follow this trust model, as user-supplied \nt{}s are application specific and cannot be shared.

%
After deploying \nt{}s, a user can specify one or more user-written or compiler-generated~\cite{clicknp-sigcomm16,NFP-sigcomm17} DAGs of the deployed \nt{}s. Different from traditional NT execution flows that execute NTs in sequence, we also allow multiple \nt{}s to execute in parallel. 
The \snic\ stores user-specified DAGs in its memory and assigns a unique identifier (UID) to each DAG.
At run time, each packet carries a UID, which \snic\ uses to fetch the DAG.

Finally, in addition to \nt\ DAGs, user also supply their desired ingress bandwidth for each \nt\ DAG.
In a cloud setting, this desired ingress bandwidth could be viewed in the same way as how clouds today ask users to specify the size of a VM.
Our fairness algorithm will guarantee that all users get at least their desired ingress bandwidth (\S~\ref{sec:policy}).
%

\subsection{Board Architecture and Packet Flow}

Figure~\ref{fig-board} illustrates the high-level architecture of the \snic\ board.
\snic's data plane handles all packet processing. It consists of reconfigurable hardware (FPGA) for running \nt{}s (blue parts in Figure~\ref{fig-board})
and a small amount of non-reconfigurable hardware (ASIC) for non-\nt\ systems stacks.
%
\snic's control plane is responsible for setting up policies and scheduling \nt{}s. It runs as software on a small set of general-purpose cores (SoftCores for short) (\eg, a small ARM-based SoC).
Although by design, \snic\ consists of an FPGA, an ASIC, and SoftCores, in our prototype, we built everything on FPGA for ease of deployment. 

{
\begin{figure*}[th]
\begin{minipage}{1.1\columnwidth}
\begin{center}
\centerline{\includegraphics[width=0.9\textwidth]{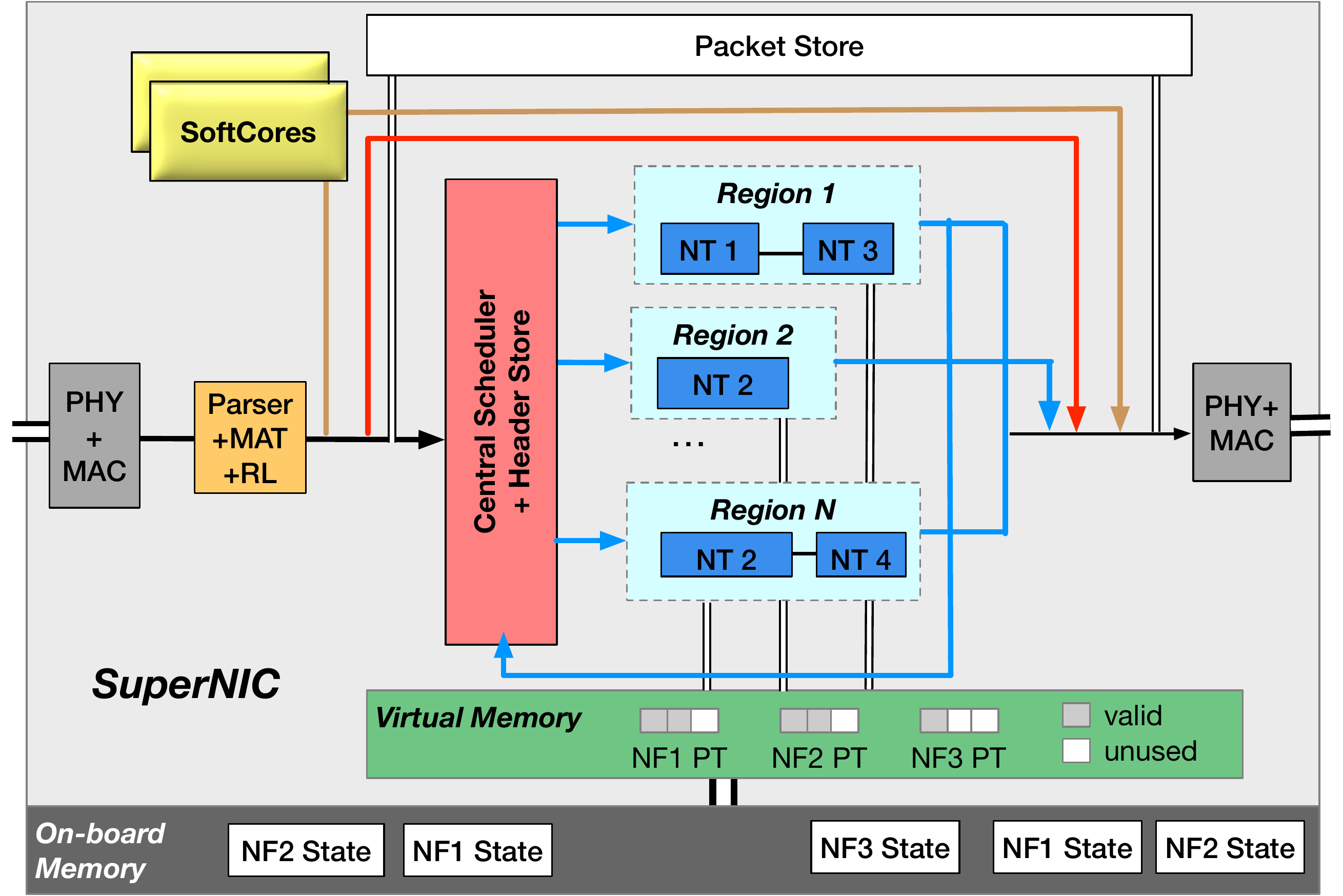}}
\vspace{-0.1in}
\mycaption{fig-board}{\snic\ On-Board Design.}
{
RL: Rate Limiter. PT: Page Table.
Orange lines: control message path.
Red lines: packets with no NT processing.
}
\end{center}
\end{minipage}
\begin{minipage}{0.05in}
\hspace{0.05in}
\end{minipage}
\begin{minipage}{0.9\columnwidth}
\hspace{0.2in} \includegraphics[width=0.8\textwidth]{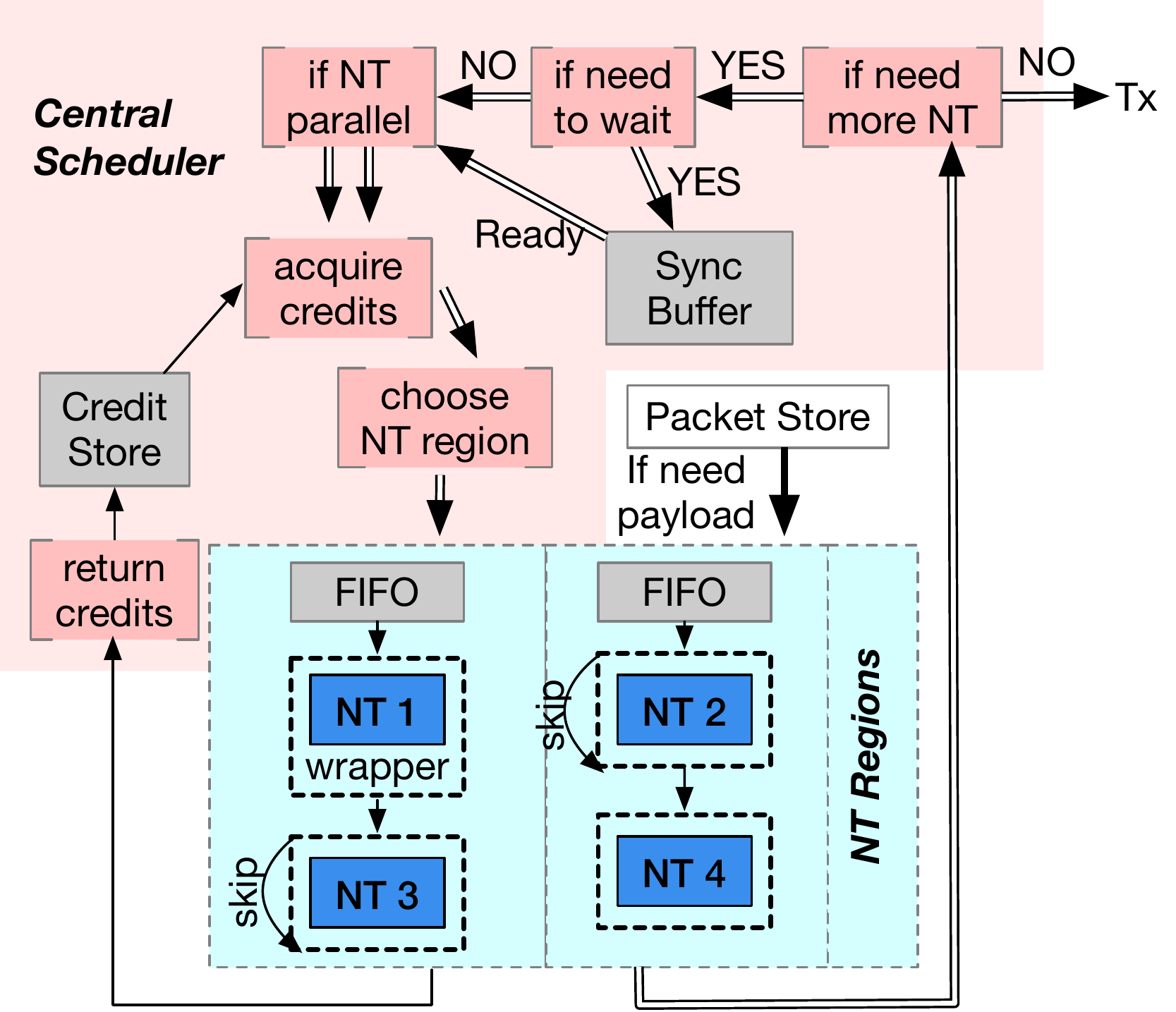}
\vspace{-0.1in}
\mycaption{fig-sched}{\snic\ Packet Scheduler and \nt\ Region Design.}
{
Double arrows, single arrows, and thick arrows represent packet headers, credits, and packet payload.
}
\end{minipage}
\end{figure*}
}

When a packet arrives at an RX port, it goes through a standard physical and reliable link layer.
Then our parser parses the packet's header  and uses a Match-and-Action Table (MAT) to decide where to route the packet next.
The parser creates a packet descriptor for each packet and attaches it to its header. The descriptor contains fields for storing metadata, such as an \nt\ DAG UID.
Data-plane packet payloads are sent to the {\em packet store}. Their headers go to a central scheduler. 
The scheduler determines when and which \nt(s) will serve a packet and sends the packet accordingly.
After an \nt\ chain finishes, if there are more \nt{}s to be executed, the packet is sent back to the scheduler to begin another round of scheduling.
When all \nt{}s are done, the packet is sent to the TX port.

\section{SuperNIC Design}
\label{sec:design}

A key and unique challenge in designing \snic{}s is space- and performance-efficient execution of hardware-based \nt{}s in a multi-tenant environment.
Moreover, we target a dynamic environment where not only the load of an application but also the applications themselves could change from time to time.
Thus, unlike traditional SmartNICs that focus on packet processing and packet scheduling, \snic\ also needs to schedule \nt{}s efficiently.
We design \snic\ to simultaneously achieve several critical goals:

\begin{itemize}

\myitem{(G1)} a system stack (non-\nt\ parts) that can process packets at line rate. 

\myitem{(G2)} high-throughput, low-latency \nt\ DAG execution.

\myitem{(G3)} quick adaptation to workload changes.

\myitem{(G4)} efficient usage of on-board hardware resources. 

\myitem{(G5)} safe and fair sharing of all on-board resources.

\end{itemize}

This section first discusses how \snic\ organizes, deploys, launches, auto-scales, and parallelizes \nt{}s. We then discuss how we schedule packets and how we ensure fairness, and finally, how we build our virtual memory system.

\subsection{\nt\ Region and \nt\ Chain}
\label{sec:nt}

\bolditpara{\nt\ chains and \nt\ regions.}~~
As more tenants occupy a server and more types of workloads start to benefit from network task offloading, we anticipate \snic\ to handle not only more \nt{}s but also more complex {\em DAGs of \nt{}} than a normal SmartNIC.
The first challenge that arises from this is the on-board connectivity between the \nt{}s and the scheduler.
Previous solutions like PANIC~\cite{panic-osdi20} connect all \nt{}s and the scheduler to a central crossbar. The area and performance overheads of a crossbar grow dramatically as its connectivity increases.
To tackle this problem, \novelty{we propose to chain \nt{}s and put one chain (\ie, a sequential list of \nt{}s) in one {\em \nt\ region}}, as shown in Figure~\ref{fig-board}. Our observation is that applications usually have the same processing chain for all or most flows, and the chain can thus be accessed as one unit. We then use a much smaller crossbar to connect the central scheduler to the \nt\ regions, thereby reducing hardware complexity and area cost (\textbf{G4}).
One \nt\ DAG can have multiple parallel chains in multiple regions, but we do not put parallel chains in one region, as parallel execution of chains need scheduler involvement (\S\ref{sec:parallel}). 

An implication of the above \nt\ chain design is that each unique chain requires its own region, even when some \nt{}s in the chain are the same as \nt{}s in other chains.
To more efficiently utilize FPGA space (\textbf{G4}), \novelty{we propose a mechanism for re-using a part of a chain among multiple \nt\ DAGs. Our idea is to allow the {\em skipping} of arbitrary \nt(s) in a chain (Figure~\ref{fig-sched})}.
For example, 
in Figure~\ref{fig-pipeline}, when deploying DAGb, \snic\ can directly use \nt{}2 and \nt{}4 in deployed chains of DAGa by skipping \nt{}1 and \nt{}3 in these chains.

\bolditpara{\nt\ region.}~~
An {\em \nt\ region} is the unit to launch an \nt\ chain.
Each region can be independently re-programmed via FPGA {\em partial reconfiguration (PR)}~\cite{pr-fpl09}.
Since the FPGA areas for PR need to be pre-determined before launching the FPGA, the region size also needs to be pre-configured.
A larger region could fit more \nt{}s in a chain but would waste FPGA space when \nt{}s cannot fill the whole region.
Our advised heuristic is to choose a relatively small region size because smaller regions enable more efficient usage of FPGA space. 
A tradeoff of smaller regions is that \nt{} chains that occupy more space will need to be broken up into sub-chains in multiple regions, and between sub-chains a packet needs to go back to the scheduler.
However, a smaller region could still host long chains of {\em small \nt{}}s. We observe that it is more beneficial to chain smaller \nt{}s together, as smaller \nt{}s are likely to run shorter. The alternative of not chaining them and going through the scheduler after each small \nt\ would have a bigger relative performance impact on it. Thus, a smaller region size achieves both efficient space utilization and good application performance.
It is also possible to divide an FPGA into regions of different sizes, as shown by previous work~\cite{amorphos-osdi18,vital-asplos20}.
In \snic's context, regions of different sizes could complicate the offline DAG compilation process and the fairness algorithm.
We leave this exploration to future work.

\subsection{\nt\ Pipelining and Parallelism}
\label{sec:parallel}

When executing an NT DAG, we exploit various pipelining and parallelism, as illustrated in Figure~\ref{fig-pipeline}.
First, we pipeline a chain of NTs by dividing it into individual NT stages and sending a new packet to these stages in every new time cycle, as in S1 of Figure~\ref{fig-pipeline}. Here, the two DAGs share the single chain, and to execute DAGb, NT1 and NT3 are skipped.
Second, we execute different chains of a DAG in parallel to reduce the total time needed to process a packet (we call it {\em DAG parallelism}).
The \snic\ scheduler sends the {\em same packet} to each of the parallel regions by duplicating the packet header.
For example, to reduce the execution time of DAGa, we can run \nt{}1$\xrightarrow[]{}$\nt{}2 and \nt{}3 in parallel, as in S2, which reduces DAGa's execution time from four units to three units. Note that after executing NT3, the packet needs to go back to the scheduler, which waits for the completion of \nt{}1$\xrightarrow[]{}$\nt{}2 before executing \nt{}4 (\S\ref{sec:packetsched}).
Finally, we create multiple instances of the same NT chain to further increase packet-processing throughput (we call it {\em instance parallelism}).
The scheduler sends different packets in a round-robin way to the parallel instances of an \nt\ DAG (\S\ref{sec:packetsched}).
For example, we create two instances of \nt{}1$\xrightarrow[]{}$\nt{}2 and two instances of \nt{}4 in S3 to improve S2's overall throughput.

We decide the amount of DAG and instance parallelism dynamically. Based on request load to an \nt\ DAG and the fair share we assign to the user, \snic\ automatically scales the number of instances of an \nt\ DAG or a subset of it.
The number here is not necessarily an integer and can be less than one, as an \nt\ can be shared by multiple users (\eg, NT3 and NT4 in Figure~\ref{fig-pipeline}). For DAG-level parallelism, we infer the \nt{}s that can run in parallel within a DAG from the DAG architecture and execute them in different regions if the user's fair share of FPGA regions allow and if doing so could reduce the DAG's execution time.
We defer the discussion of policy for determining the amount of these parallelism to \S\ref{sec:policy}.

{
\begin{figure}[th]
\begin{center}
\centerline{\includegraphics[width=\columnwidth]{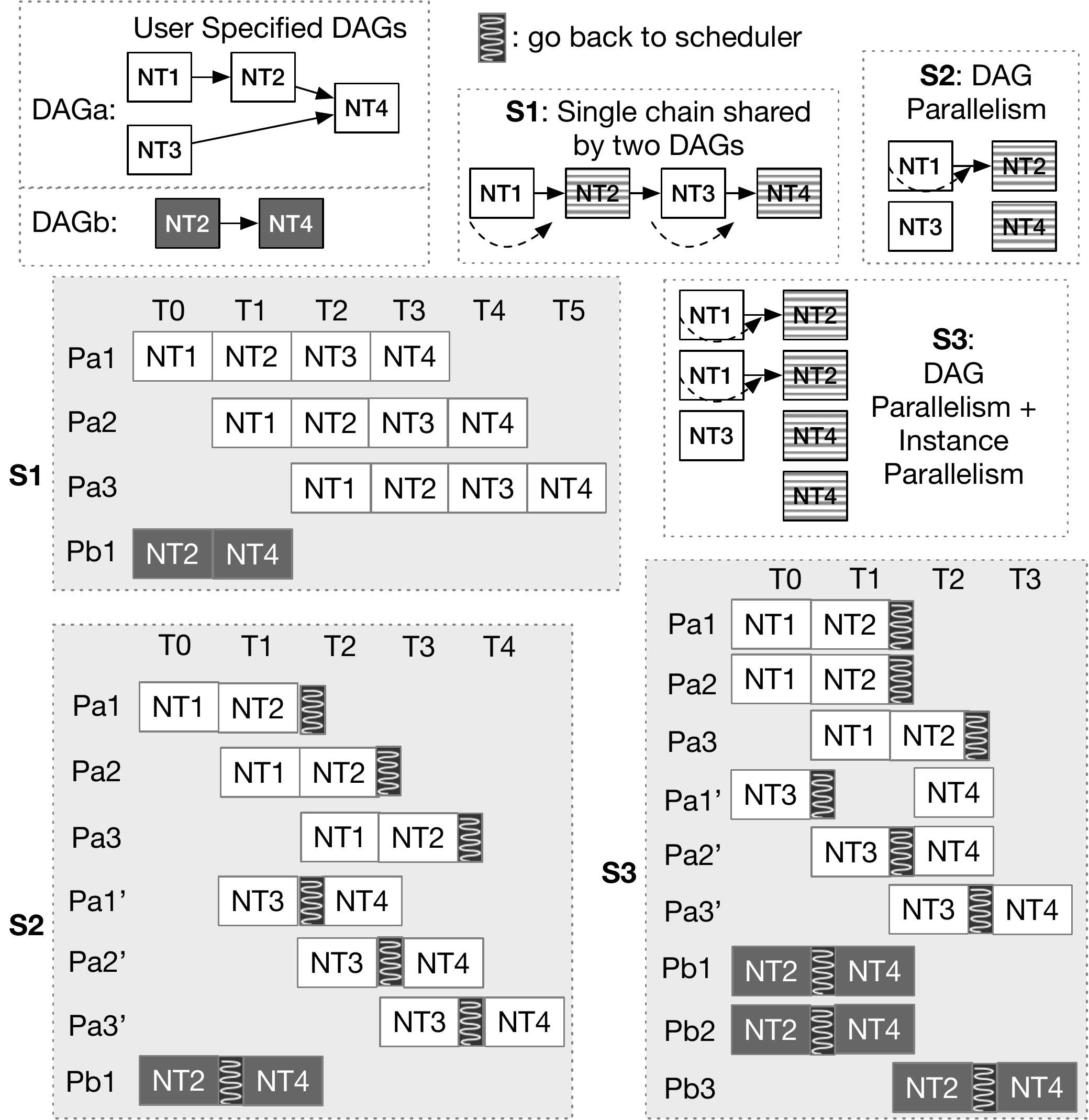}}
\vspace{-0.1in}
\mycaption{fig-pipeline}{\snic\ NT Pipeline.}
{
Two deployed DAGs, a and b. S1, S2, and S3 are three ways of executing them. $Pa_{i}$/$Pb_{i}$ refer to the $i$th packet targeting the first/second DAG, $P_{i}'$ refers to a forked packet. $T_{i}$ refers to a time unit in the timeline.
}
\end{center}
\vspace{-0.1in}
\end{figure}
}

\subsection{\nt\ Deployment and Launching} 
\label{sec:nt-deploy}

\bolditpara{\nt\ deployment.}~~
Users deploy \nt{}s to the \snic\ platform ahead of time as FPGA netlists and specify DAG of these \nt{}s.
We generate a set of FPGA bitstreams, each for a subset of the DAG whose size does not exceed a region (Figure~\ref{fig-nt-example}). 
We currently enumerate all possible subsets at deployment time so that \snic\ can quickly choose any subset to launch, share, or duplicate at run time without waiting for slow bitstream-generation phases. Future compiler works could optimize this step by better selecting the subsets to generate.
When generating bitstreams, we attach a small \snic\ wrapper to each \nt\ (Figure~\ref{fig-sched}).
This wrapper is essential: it enables skipping an \nt\ in a chain, monitors the runtime load of the \nt\ (\S\ref{sec:policy}), ensures signal integrity during PR, and provides a set of virtual interfaces for \nt{}s to access other board resources like on-board memory (\S\ref{sec:memory}).
We store pre-generated bitstreams in the \snic{}'s on-board memory; each bitstream is small, normally less than 5\MB.

{
\begin{figure*}
\begin{center}
\centerline{\includegraphics[width=0.95\textwidth]{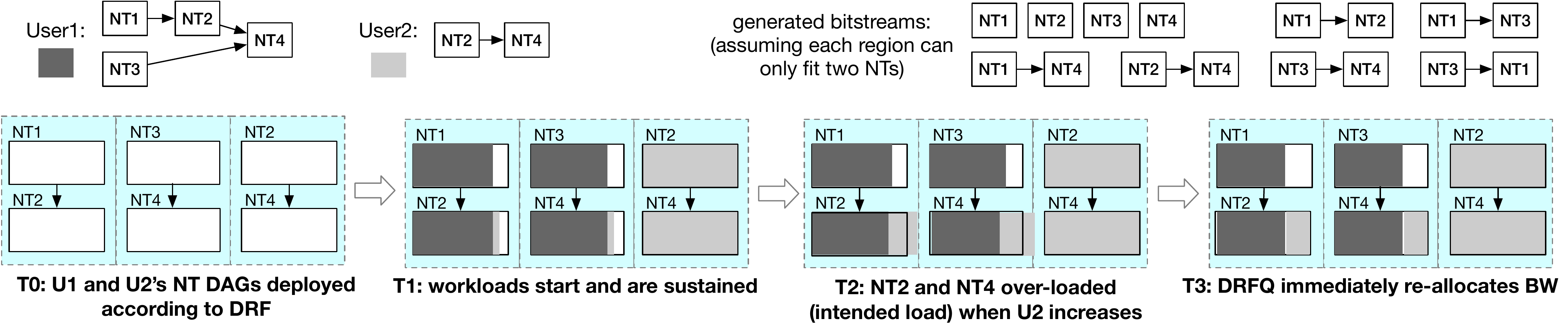}}
\vspace{-0.15in}
\mycaption{fig-nt-example}{An Example of \nt\ chaining and scheduling.}
{
Top: user1 and user2's \nt\ DAGs and \snic's generated bitstreams for them.
Bottom: timeline of \nt\ bandwidth allocation change.
Dark grey and light grey represent user1 and user2's load.
The launched chains are \nt{}1$\xrightarrow[]{}$\nt{}2, \nt{}3$\xrightarrow[]{}$\nt{}4, and \nt{}2$\xrightarrow[]{}$\nt{}4
with the first two chain's \nt{}2 and \nt{}4 being shared by both users.
The maximum throughput of each NT is 10\Gbps.
U1's requested ingress bandwidth is 8\Gbps, and U2's requested ingress bandwidth is 14\Gbps.
}
\end{center}
\vspace{-0.15in}
\end{figure*}
}

\bolditpara{\nt\ chain launching.}~~
We now discuss how we launch \nt\ chains.
Our overall strategy is to avoid FPGA partial reconfiguration (PR) or hide its cost as much as possible (\textbf{G2},  \textbf{G3}), as it is relatively slow: the maximum reported achievable PR throughput is around 800\,MB/s~\cite{coyote-osdi20}, or about 5\,\ms\ for our default region size.


Our first idea is to \novelty{{\em pre-launch} \nt{}s to avoid application-perceived PR time}.
Specifically, when the user deploys an \nt\ DAG (prior to when it is accessed), we check if any of its \nt{}s is missing on an \snic. If so, and if there are free regions on the \snic, we launch them at these regions. 


Afterwards, when a deployed \nt\ DAG is first accessed, we check if any of its \nt{}s are the same as existing \nt{}s on the \snic. 
If so, and if the existing \nt{}s still have available bandwidth, we \novelty{time share the existing \nt{}s with the new application}. 
In this case, new traffic can be served immediately.
Otherwise, we check if there is a free region.
If so, we enable space sharing by launching the \nt\ at this region, and new traffic can be served after FPGA PR finishes.

\novelty{If all of the above fail, we invoke a context switch}, which is the slowest on-board operation in \snic\ and thus our last resort. 
To perform a context switch, the SoftCore picks the region that is least loaded and goes through a {\em stop-and-launch} process. The SoftCore sends a signal to the current \nt{}s to let them ``{\em stop}''.
These \nt{}s then store their states in on-board memory to prepare for the stop.
At the same time, the SoftCore informs the scheduler to stop accepting new packets. 
The scheduler will buffer packets received after this point.
After the above steps finish, the SoftCore reads the new bitstream from the on-board memory and starts the FPGA PR process ({\em launch}).
Afterwards, the newly launched chain can start serving packets, and it will first serve previously buffered packets, if any.

To further reduce PR costs, we use a technique similar to the traditional victim cache~\cite{VictimCache-ISCA90}. We \revise{cache} a de-scheduled \nt\ chain in a region around for a while unless the region is needed to launch a new chain. If the de-scheduled \nt\ chain is accessed again during this time, we can directly use it in that region, reducing the need to do PR at that time.


\subsection{Packet Scheduling Mechanism}
\label{sec:packetsched}

We now discuss the design of \snic's packet scheduling mechanism. Figure~\ref{fig-sched} illustrates the overall flow of \snic's packet scheduling and execution.
Based on the \nt-chain architecture, we propose a scheduling mechanism that reduces scheduling overhead and increases the scalability of the scheduler (\textbf{G1}, \textbf{G2}).
\novelty{Our idea is to} \revise{try \textit{reserving}} \novelty{credits for an {\em entire} \nt\ chain in a region as much as possible and then execute the chain as a whole without involving the scheduler in the middle}

\revise{Executing chains in their entirety} improves both the packet's processing latency and the central scheduler's scalability (\textbf{G5}).
\revise{However, to reserve credits for a whole chain, the schedule needs to wait until each \nt\ in the chain has available credits, which could delay the processing of packets. To mitigate this issue, we only reserve credits for whole chains \textit{opportunistically}. If all \nt{}s in a chain have available credits when a packet is about to be scheduled, the scheduler reserves a credit from each of these \nt{}s. Otherwise, when not all \nt{}s have available credits, we fall back to PANIC~\cite{panic-osdi20}'s scheduling policy by sending the packet to each subsequent \nt\ when it has credits or to the scheduler when it does not.}
To utilize multiple instances of an \nt{} chain (\S\ref{sec:parallel}), our scheduler \revise{pipelines} different packets to the instances in a round-robin fashion.
To explore DAG parallelism, the scheduler makes copies of the packet header and sends them to these regions concurrently. To obey the order of \nt{}s that users specify, we maintain a {\em synchronization buffer} to store packet headers after they return from an \nt{} chain's execution and before they can go to the next stage of \nt{}s (Figure~\ref{fig-sched}).

\subsection{Fairness Policy}
\label{sec:policy}

As we target a multi-tenant environment, \snic\ needs to fairly allocate its resources to different users (\textbf{G5}). These resources include ingress and egress ports, packet-store buffer space, on-board memory space, and FPGA chip area. 
As explained in \S\ref{sec:related-fairness}, traditional fairness solutions that target server environments fairly allocate different portions (\ie, {\em space shares}) of each type of resource among multiple jobs, and prior fairness solutions that target network devices {\em time share} each type of resource. 

\novelty{Different from these traditional environments, we integrate both space and time sharing on an \snic\ for more efficient consolidation.
\revise{This is especially hard for FPGA\footnote{We time-share other resources like ingress/egress ports and packet buffer and space-share on-board memory.}, because we aim to partition an FPGA chip into regions that can be space shared across users and to time share a region across different \nt\ chains via context switching.
In addition, we also aim to support the time sharing of the same \nt\ chain across packets from different users. 
Finally, unlike server computing jobs, network load can vary quickly and by considerable amounts.}}

To confront the above challenge, \novelty{our proposal is to tackle space and time sharing one at a time, while ``freezing'' the other, and to dynamically adjust fair allocation based on {\em observed load}},
as explained below and illustrated in the example in Figure~\ref{fig-nt-example}.
Overall, our fairness algorithm guarantees a {\em sharing incentive} and {\em strategy-proofness} with all \snic's hardware resources.
As defined in the DRF work~\cite{DRF}, a sharing incentive means that each user should be better off sharing the resources than exclusively using her own partition of resources, and strategy-proofness means that users should not be able to benefit by lying about their resource demands.

\bolditpara{Resource requirement and load monitoring.}
For a fairness algorithm to work, it needs to know the user demand for each type of resource.
Different from previous works~\cite{DRF,DRFQ} that ask users to specify this demand for all resources, \novelty{we use a combination of methods to more automatically and precisely capture resource demands.
As introduced in \S\ref{sec:interface}, users only need to supply their desired ingress bandwidth going to each \nt\ DAG. This is the service level that \snic\ will deliver (if \snic\ finds that the requirement cannot be met, it will reject the \nt\ DAG).
\snic\ automatically interprets other demands.
When the \snic\ platform generates bitstreams of \nt\ DAGs (\S\ref{sec:nt-deploy}), it captures the FPGA area consumption of the DAG. It also estimates the memory consumption and the egress bandwidth based on the desired ingress bandwidth~\cite{DRFQ}.
}



After the user workload starts, \novelty{\snic\ keeps monitoring the load that is needed by each user at every type of resource. 
If the actual load is lower than the user-input demand, \snic\ will only allocate for the actual load, harvesting the remaining for other users.
Here, we capture the \textit{intended} load requirement, not the load that is actually handled by a board unit}.
For example, for each user, the central scheduler measures the rate of packets that should be sent to an \nt\ before assigning credits; \ie, even if there is no credit for the \nt, we still capture the intended load it should handle.
We capture the intention so that \snic\ can properly and promptly react to potentially satisfy it (under the user's fair share).

\bolditpara{Step 1: Fair space sharing.}~~
Different from traditional networking devices, \snic\ divides its FPGA area and on-board memory spatially across user \nt{}s. 
Our first step in achieving fair share is to find a fair spatial allocation and determine how many instances of an \nt\ chain (\ie, number of regions) to launch and how much on-board memory to assign to each user.
%
\novelty{
Different from traditional server settings where each resource has a maximum space and users specify the space they need within each resource~\cite{DRF}, our users instead specify their required bandwidth.
Thus, we represent the demanded FPGA size to be the FPGA area multiplied by the ratio of user-required bandwidth to the \nt{}'s maximum bandwith.}
For example, if a user requires 10\Gbps\ and uses an \nt\ DAG that occupies 4 units of FPGA area and supports a maximum of 40\Gbps, we will calculate the user's demand for FPGA as 1 unit ($4 * 10/40$). 

With this representation, we then calculate the FPGA area and memory space to allocate to each user by solving a set of linear equations as in DRF (\S\ref{sec:related-fairness}).
\novelty{The decision may result in the launch of multiple instances of a DAG (assignment greater than one region) or multiple users sharing \nt{}s (assignment less than one region or the non-integer part of an assignment).}
Afterwards, we adjust the FPGA area and on-board memory allocation accordingly.
We trigger this spatial-allocation step when a new \nt{} DAG is launched or an existing \nt\ DAG is de-scheduled. Thus, even though spatial allocation adjustment involves the slow PR process, it only happens infrequently.

\bolditpara{Step 2: Fair time sharing.}~~
Between two space-sharing steps, the space allocation does not change. 
\novelty{Thus, we can treat each \nt\ DAG as a fixed-size resource type, reducing the time-sharing problem to a traditional one. 
For the time-share allocation, we use the monitored load, since, unlike the space-sharing phase, we can now perform a more fine-grained allocation of time without the need to perform PR.}
We run the DRFQ~\cite{DRFQ} algorithm 
to assign a virtual start time and virtual finish time to each packet, allowing users to time share each \nt\ region, the ingress bandwidth, the egress bandwidth, and the packet store buffer, based on the monitored load at each resource type for each user. 
Note that we do not time-share on-board memory,
since it is usually used to store longer-term data that lives beyond a packet.

Although the calculations in Step 1 and Step 2 are each similar to DRF and DRFQ, we apply them in a novel way: we first treat an FPGA as a single resource with capacity calculated in consideration of area and bandwidth; we then treat each NT region as a separate type of resource and fairly time share them.
Another major difference is how we provide strong upper-bound guarantees on the utilization of each resource type for each user. 
\novelty{Instead of throttling a user's packets at each \nt\ and every type of resource to match the desired allocation, we only control the user's ingress bandwidth allocation.
Our observation is that since each \nt's throughput for an application, its packet buffer space consumption, and egress bandwidth are all proportional to its ingress bandwidth, we could effectively control these allocations through the ingress bandwidth allocation.
Doing so avoids the complexity of throttling management at every type of resource.
Moreover, throttling traffic early on at the ingress ports helps reduce the load going to the central scheduler and the amount of payload going to the packet store.}

\bolditpara{An example.}~~
We now illustrate how our fairness algorithm works with a simple example in Figure~\ref{fig-nt-example}.
There are two users running two DAGs on an \snic\ with 6 units of FPGA resource (\eg, LUTs and BRAM). The FPGA is divided into three regions, each of 2 units and each sustaining 10\Gbps.
Each \nt\ occupies 1 unit of FPGA resource and sustains the peak bandwidth of 10\Gbps.
We assume that U1's requested ingress bandwidth is 8\Gbps, and U2's requested ingress bandwidth is 14\Gbps.
We further assume that U1's memory consumption is 1\GB, U2's memory consumption is 2\GB, and the total memory size is 10\GB.
Thus, U1's FPGA demand is $8\times4=32$ (\ie, 8/15 of the total FPGA area-bandwidth product), and U2's FPGA demand is $14\times2=28$ (\ie, 7/15 of the total). 
U1's memory demand is 1/10, and U2's memory demand is 1/5.
Thus, for both U1 and U2, their dominant resource type is FPGA.
We then run DRF with these demands, which yields the FPGA allocation of U1 to be 32 units and U2 to be 28 units. These allocations represent the bandwidth-area product assigned to each user (\textbf{T0}).
Since each region can host 20 units of bandwidth-area product, we deploy U1's DAG to two regions and U2's DAG to the third region, while letting U2 share U1's NT2 and NT4.
After the traffic starts, initially \snic\ can sustain the load of U1 (8\Gbps) and U2 (11\Gbps) (\textbf{T1}).
When U2's load increases (to 13\Gbps), \snic\ monitors that the intended load on NT2 and NT4 is 11\Gbps\ each.
\snic\ then runs DRFQ to adjust the bandwidth allocation of NT2 and NT4, according to the users' initial required load ratio (\ie, $32:28$ or $8:7$). This results the allocation of NT2/NT4 to U1 and U2 to be 7.27 and 2.73.

\subsection{Virtual Memory System}
\label{sec:memory}
\snic's allow \nt{}s to use off-chip, on-board memory.
To isolate different applications' memory spaces and to allow the over-subscription of physical memory space in an \snic, we build a simple page-based virtual memory system.
\nt{}s access on-board memory via a virtual memory interface,
where each \nt\ has its own virtual address space.
Our virtual memory system translates virtual memory addresses into physical ones and checks access permissions with a single-level page table.
Since on-board memory of network devices is usually much smaller than server memory (our test board has 10\GB) and most hardware computation only uses small amounts of memory, we make the virtual address space 1\GB\ by default.
We use huge pages (2\MB\ size) to reduce the amount of on-chip memory to store the page table.
With these configurations, we simply use a flat (one-layer) page table, which amounts to 4\KB\ per address space.
Physical pages are allocated on demand; when a virtual page is first accessed, \snic\ allocates a physical page from a free list.



\section{Evaluation Results}
\label{sec:results}

We implemented \snic\ on the HiTech Global HTG-9200 board~\cite{htg9200}, which has nine 100\Gbps\ ports, 10\GB\ on-board memory, and a Xilinx VU9P chip with 2,586K LUTs and 43\MB\ BRAM.
We implemented most of \snic's data path in SpinalHDL~\cite{spinalhdl} and \snic's control path in C (running in a MicroBlaze SoftCore~\cite{microblaze-xilinx} on the FPGA).
Most data path modules run at 250 MHz.
In total, \snic\ consists of 23.8k SLOC.
Figure~\ref{fig-fpga-resource} shows the FPGA resource consumption of different modules in \snic\ \revise{and our implemented NTs}.
The core \snic\ modules consume less than 5\% resources of the Xilinx VU9P chip, leaving most of it for \nt{}s.

\bolditpara{Environment.}~~ 
We perform both cycle-accurate simulation (with Verilator~\cite{verilator-site}) and real end-to-end deployment.
For the end-to-end deployment, we use a cluster with a 100\Gbps\ Ethernet switch, an HTG-9200 board, two Dell PowerEdge R740 servers each equipped with a Xeon Gold 5128 CPU and an NVidia 100\Gbps\ ConnectX-4 NIC, and a Xilinx 10\Gbps\ ZCU106 board running as the Clio~\cite{clio-arxiv} disaggregated memory device.
\revise{
We also project sNIC's latency in a potential ASIC implementation in a similar way as previous work~\cite{Clio}. We collect the latency breakdown of time spent in third-party IPs and cycles spent in \snic\ components. We then scale the frequency of \snic\ component to 2 GHz while maintaining the amount of time spent in third-party IPs. This estimate is conservative as most of the latency is introduced in the third-party MAC and PHY modules. Real ASIC implementations of these IPs would lower overall latency further.
}

For most experiments, we use PANIC as a baseline. As PANIC's open-source code is specific to their FPGA setup, we re-implemented PANIC's core scheduling mechanism on our FPGA platform. It uses the same other on-board components like MAC and PHY as \snic.
Our baseline is our own implementation of PANIC's scheduling mechanism on our platform, where everything else is the same as \snic.

\subsection{\nt{} and \nt{} DAG Implementation} 
\label{sec:application}

We implemented three types of \nt{}s and constructed several realistic \nt{} DAGs, which we will present now.

\bolditpara{Transport and traditional NTs.}~~
To demonstrate \snic's ability of supporting transport offloading, we implemented a simple reliable transport using the go-back-N protocol on top of a lossless network.
When the receiver receives an out-of-order packet, it simply discards it and sends a NACK to the sender.
When the sender sees a NACK, it will retransmit all packets that were sent after the last acknowledged packet.

We then implement a set of \nt{}s that represent what cloud users use in a Virtual Private Cloud (VPC) setting. VPC allows users to have an isolated network environment where their traffic is not affected by others and where they can deploy their own network functions such as firewall and encryption.
We implemented four \nt{}s on \snic\ for VPC: network address translation (NAT), firewall, AES encryption, and load balancer.

{
\begin{figure}
\begin{center}
\centerline{\includegraphics[width=2.5in]{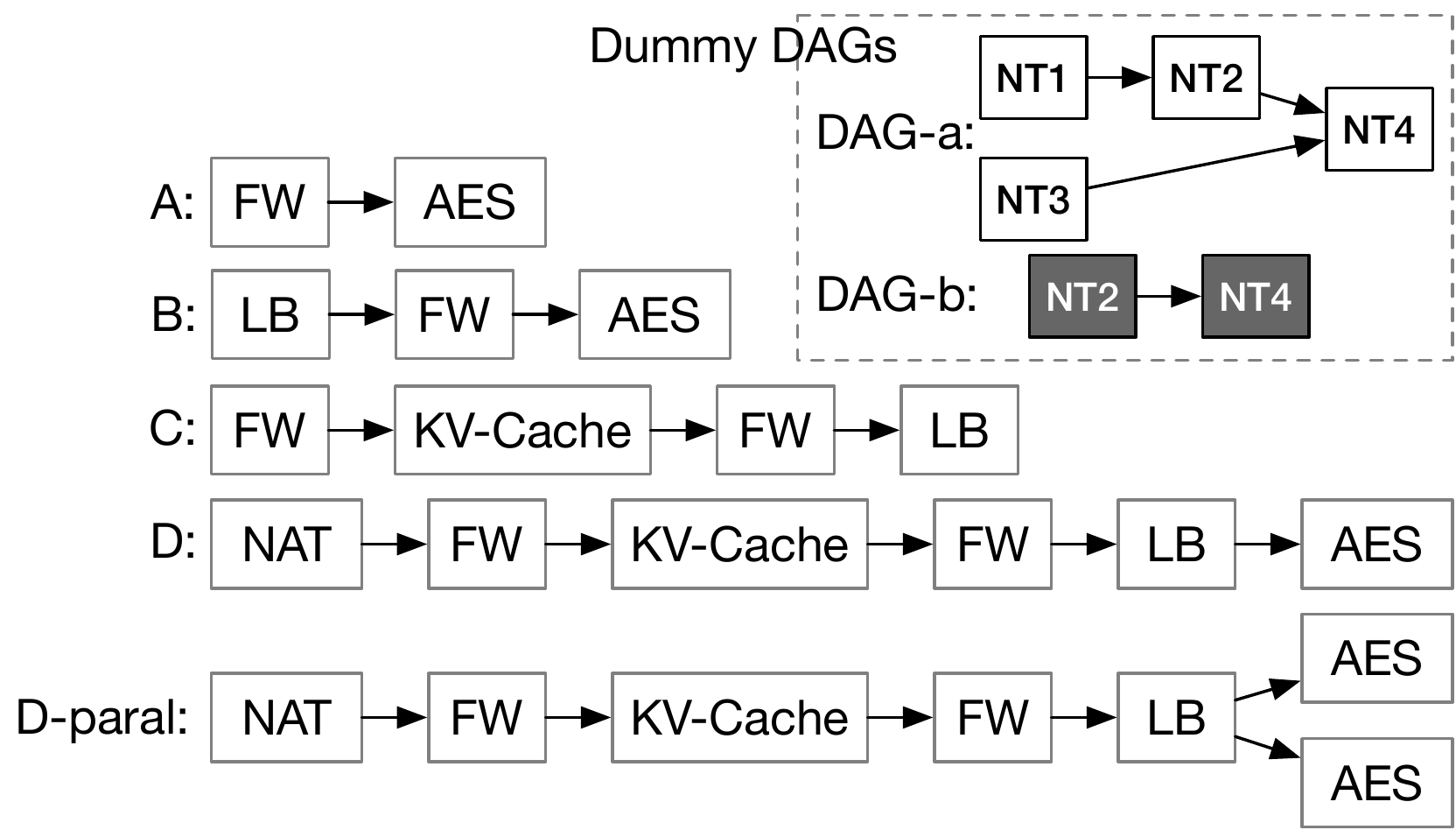}}
\vspace{-0.15in}
\mycaption{fig-nt-dag}{\nt{} DAGs Implemented.}
{
LB: Load Balancer; FW: Firewall.
}
\end{center}
\vspace{-0.3in}
\end{figure}
} 
{
\begin{figure*}[th]
\begin{minipage}{\figWidthSix}
\begin{center}
\scriptsize
\begin{tabular}{ p{0.6in} | p{0.3in} |p{0.37in} }

 \textbf{Module} & \textbf{LUT} & \textbf{BRAM} \\
\hline
\hline
\snic{} Core & 51.5K   & 102 \\
Packet Store & 10.8K   & 198 \\
PHY+MAC      & 8.5K  & 8 \\
DDR4Controller & 18.5K   & 6 \\
{Default Region} & {20K} & {40} \\
\hline
{Go-back-N} & {5359} & {0} \\
{FW/NAT} & {468/864} & {0} \\
{KV Rep/Cache} & {458/2452} & {22/11} \\
{LB} & {4533} & {0} \\

\end{tabular}
\vspace{-0.15in}
\mycaption{fig-fpga-resource}{FPGA Utilization.}
{
}
\end{center}
\end{minipage}
\begin{minipage}{1.75in}
\begin{center}
\centerline{\includegraphics[width=\columnwidth]{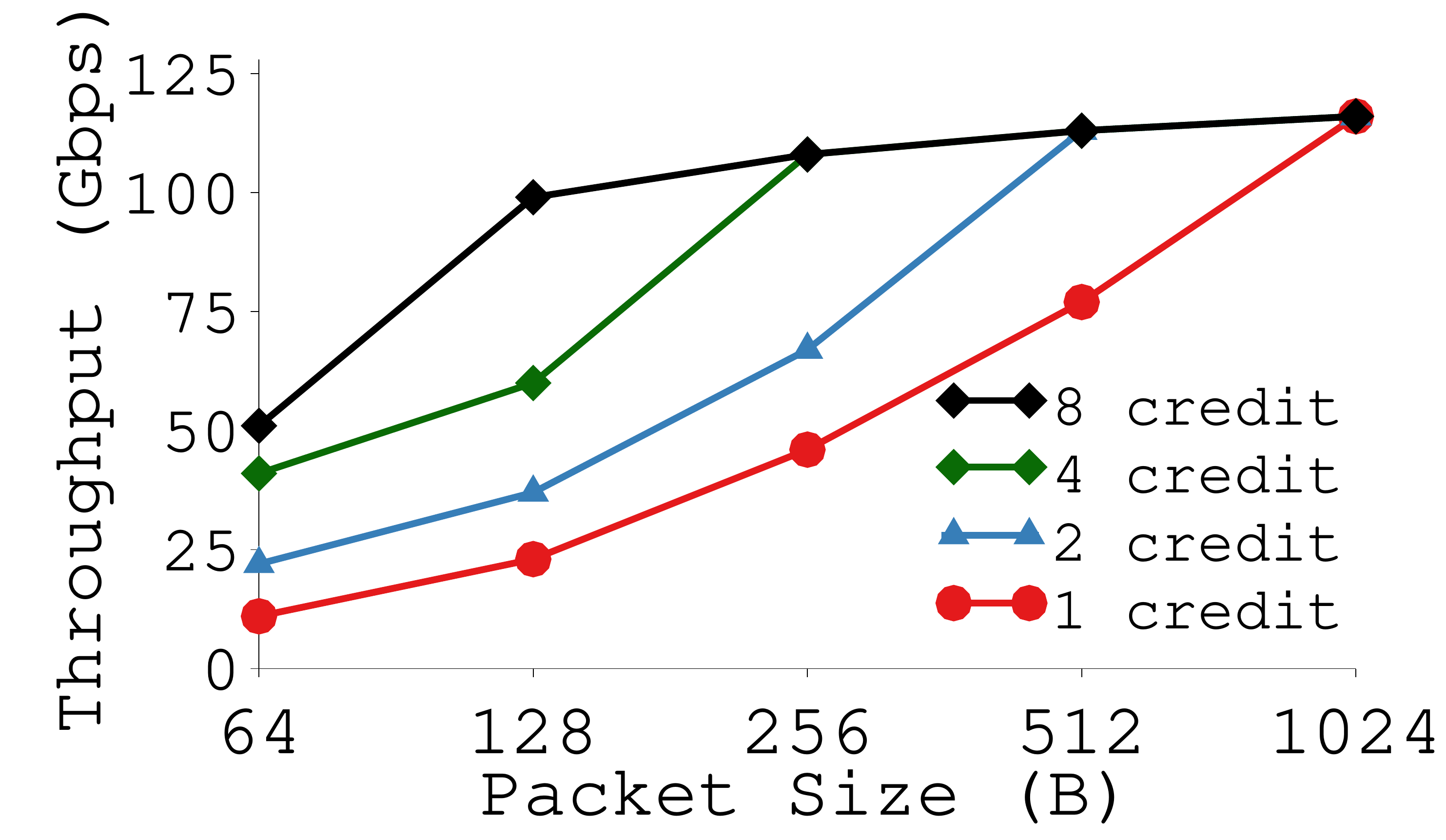}}
\vspace{-0.1in}
\mycaption{fig-credit}{Throughput with different credits.}
{
}
\end{center}
\end{minipage}
\begin{minipage}{1.57in}
\begin{center}
\centerline{\includegraphics[width=\columnwidth]{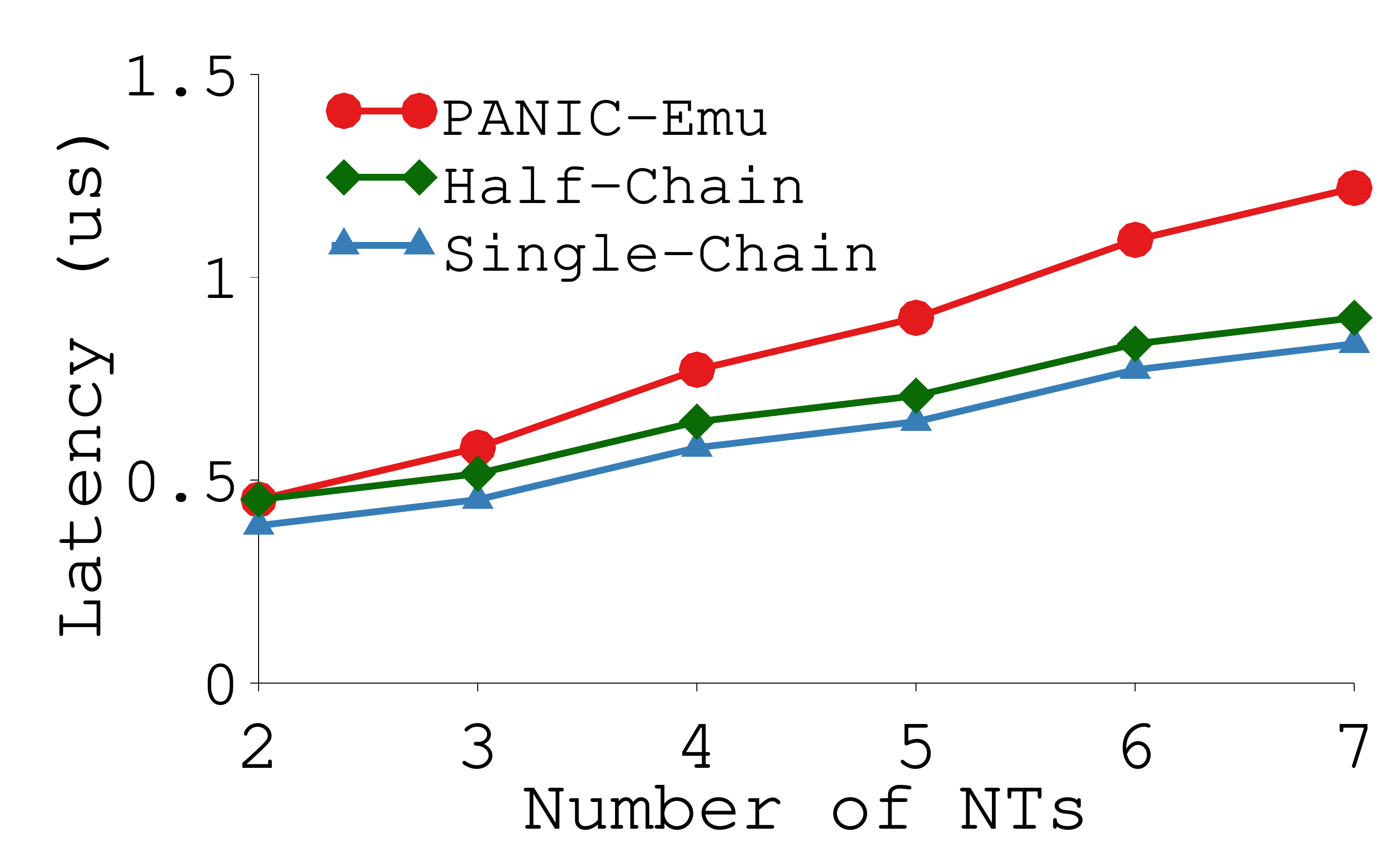}}
\vspace{-0.1in}
\mycaption{fig-nt-chain}{Dummy \nt{} Chain Latency.}
{
}
\end{center}
\end{minipage}
\begin{minipage}{1.75in}
\begin{center}
\centerline{\includegraphics[width=\columnwidth]{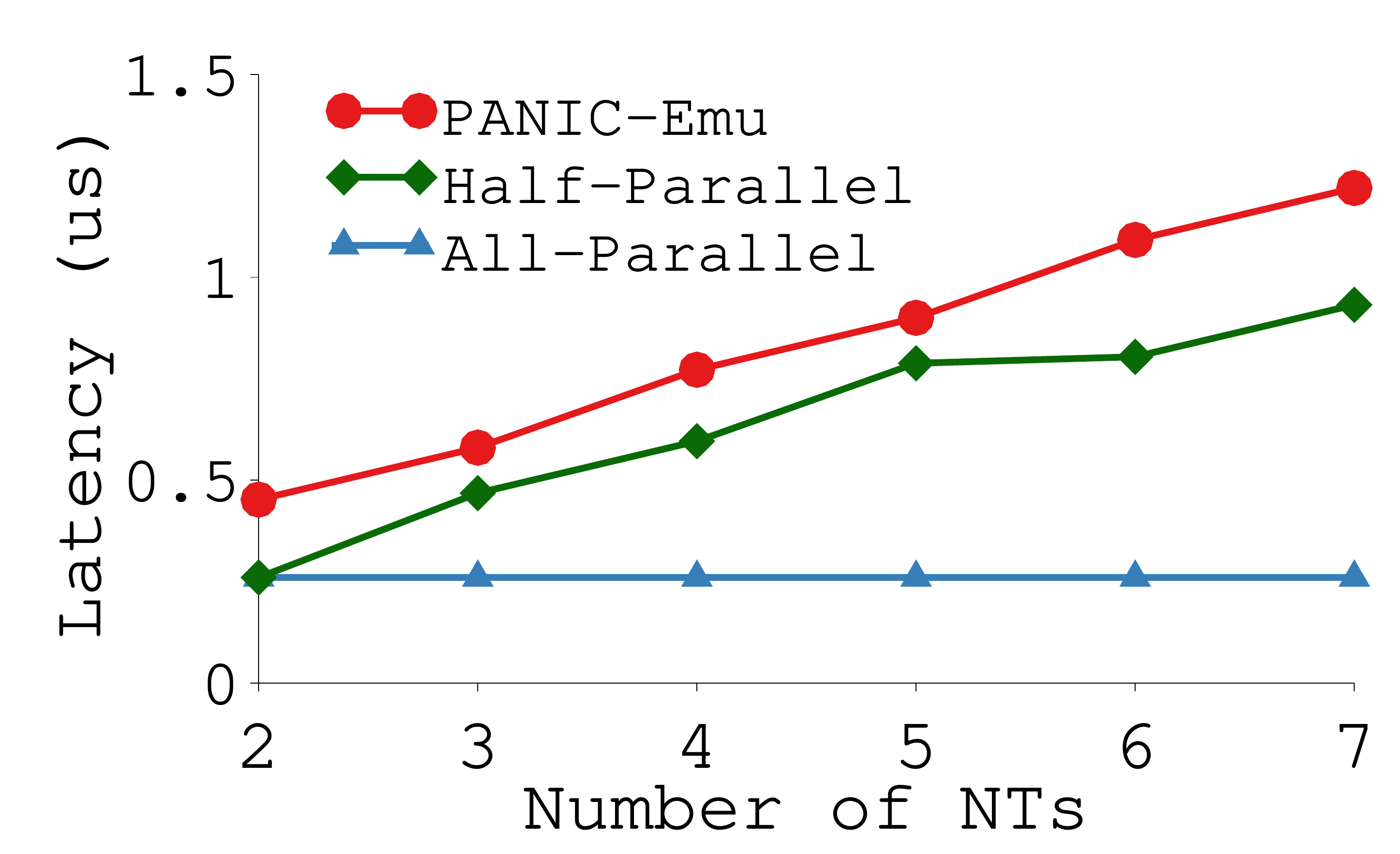}}
\vspace{-0.1in}
\mycaption{fig-nt-parallel}{Dummy \nt{} Parallelism.}
{
}
\end{center}
\end{minipage}
\vspace{-0.15in}
\end{figure*}
}

\bolditpara{Application-specific NTs.}~~
To demonstrate how users can offload application-specific tasks to \snic, we build an \nt{} for key-value stores.
The \nt\ performs key-value pair caching, where the \nt\ maintains recently written/read key-value pairs in a small buffer. If there is a cache hit, the \nt\ directly returns the value to the client. Our current implementation that uses simple FIFO replacement already yields good results. Future improvements like LRU could perform even better.


\bolditpara{\nt\ DAGs and \nt\ sharing.}~~
We deploy several \nt\ DAGs as illustrated in Figure~\ref{fig-nt-dag}. 
DAG-D is adapted from a prior network-function-chaining work~\cite{nfchain-ANCS18}.
We further deploy several \nt\ sharing and skipping cases. Here, we assume a scenario where DAG-D already runs on an \snic, and one of the DAGs, A, B, and C, is then triggered. 
Because DAG-A/B/C's NTs all exist in the DAG-D, we can leverage sharing to execute them without launching new DAGs. 
For example, to execute DAG-A, \snic\ skips NAT in DAG-D, execute FW, skips the next three \nt{}s, and finally executes AES. 

As the above real \nt\ DAGs were developed in a traditional data-center network setting that does not have the support like \snic, they are simple chains of \nt{}s. To explore more complex DAGs, we also evaluated two DAGs with dummy \nt{}s, following the example in Figure~\ref{fig-pipeline}.

\subsection{Overall Performance}
\label{sec:eval-overview}

We first test the throughput an \snic\ board can achieve with a dummy \nt.
These packets go through every functional module of the \snic, including the central scheduler and the packet store. 
We change the number of initial credits and packet size to evaluate their effect on throughput, as shown in Figure~\ref{fig-credit}.
These results demonstrate that our FPGA prototype of \snic\ could reach more than 100\Gbps\ throughput per port, and with more credits in the system, \snic\ can reach full bandwidth with smaller packet sizes.
With higher frequency, future ASIC implementation could reach even higher throughput.

Next, we evaluate the latency overhead an \snic\ FPGA board adds.  
It takes 1.3\mus\ for a packet to traverse the entire sNIC data path. 
Most of the latency is introduced by the third-party PHY and MAC modules, which could potentially be improved with ASIC implementation and/or a PCIe link. 
The \snic\ core only takes 196\,ns \revise{(or 25\,ns with ASIC projection)}.
Our scheduler achieves a small, fixed delay of 16 cycles, or 64\,ns with the FPGA frequency \revise{(8\,ns with ASIC projection)}. 
To put things into perspective, commodity switch's latency is $\mathtt{\sim}$0.8 to 1\mus.

\subsection{Deep Dive into \snic\ Designs}
\label{sec:deepdive}

We now perform a set of experiments to understand the implications of \snic's various designs. For these experiments, we generate traffic load using the Facebook distribution~\cite{fb-trace-sigcomm15}, which captures various traffic in the Facebook datacenter.

{
\begin{figure*}[th]
\begin{minipage}{1.75in}
\begin{center}
\centerline{\includegraphics[width=\columnwidth]{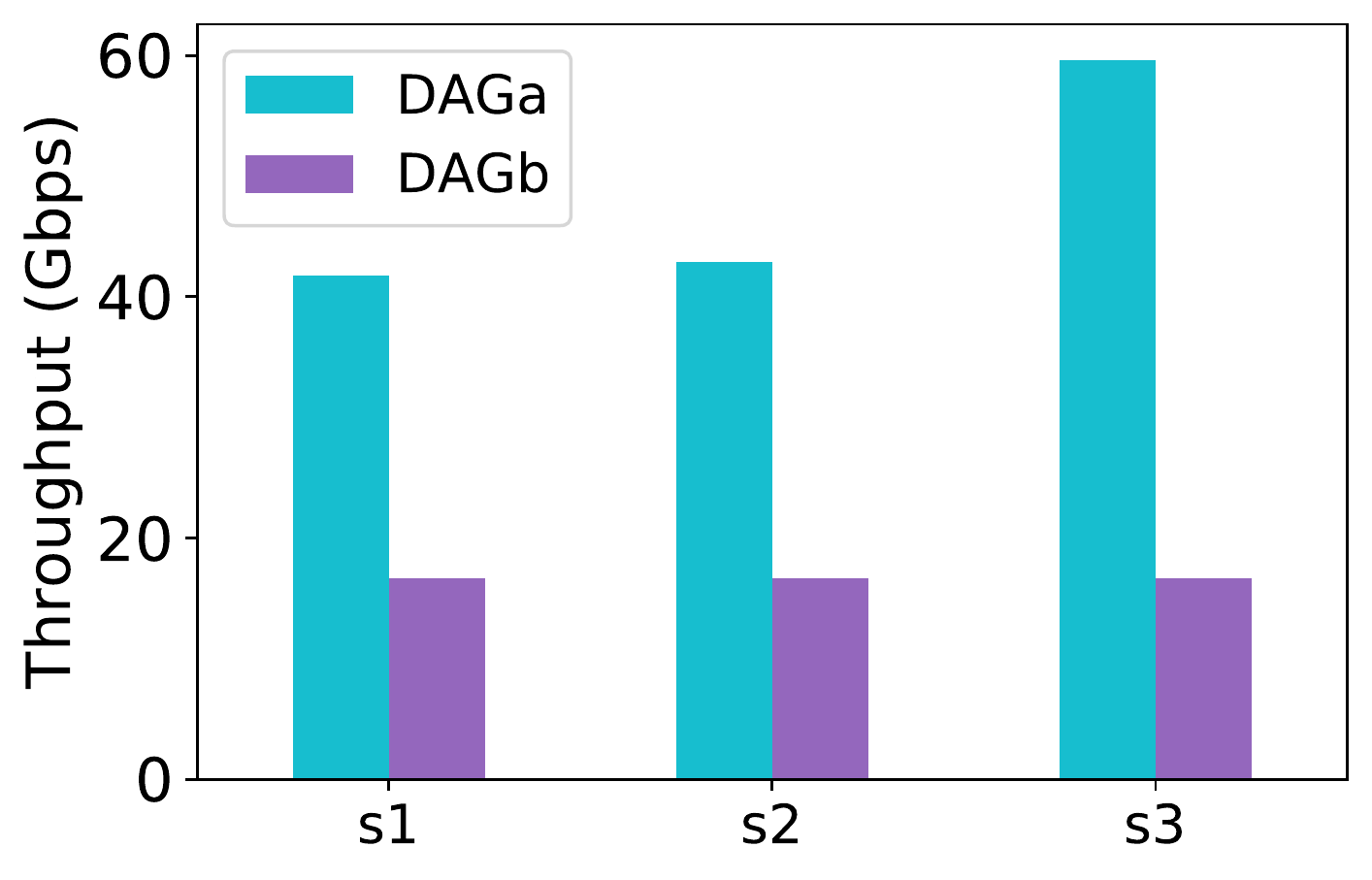}}
\vspace{-0.13in}
\mycaption{fig-dummy-nt-dag-tput}{\nt{} DAG Throughput.}
{
S1: sequential. S2: DAG parallel. S3: DAG+instance parallel (Fig.3)
}
\end{center}
\end{minipage}
\begin{minipage}{1.75in}
\begin{center}\centerline{\includegraphics[width=\columnwidth]{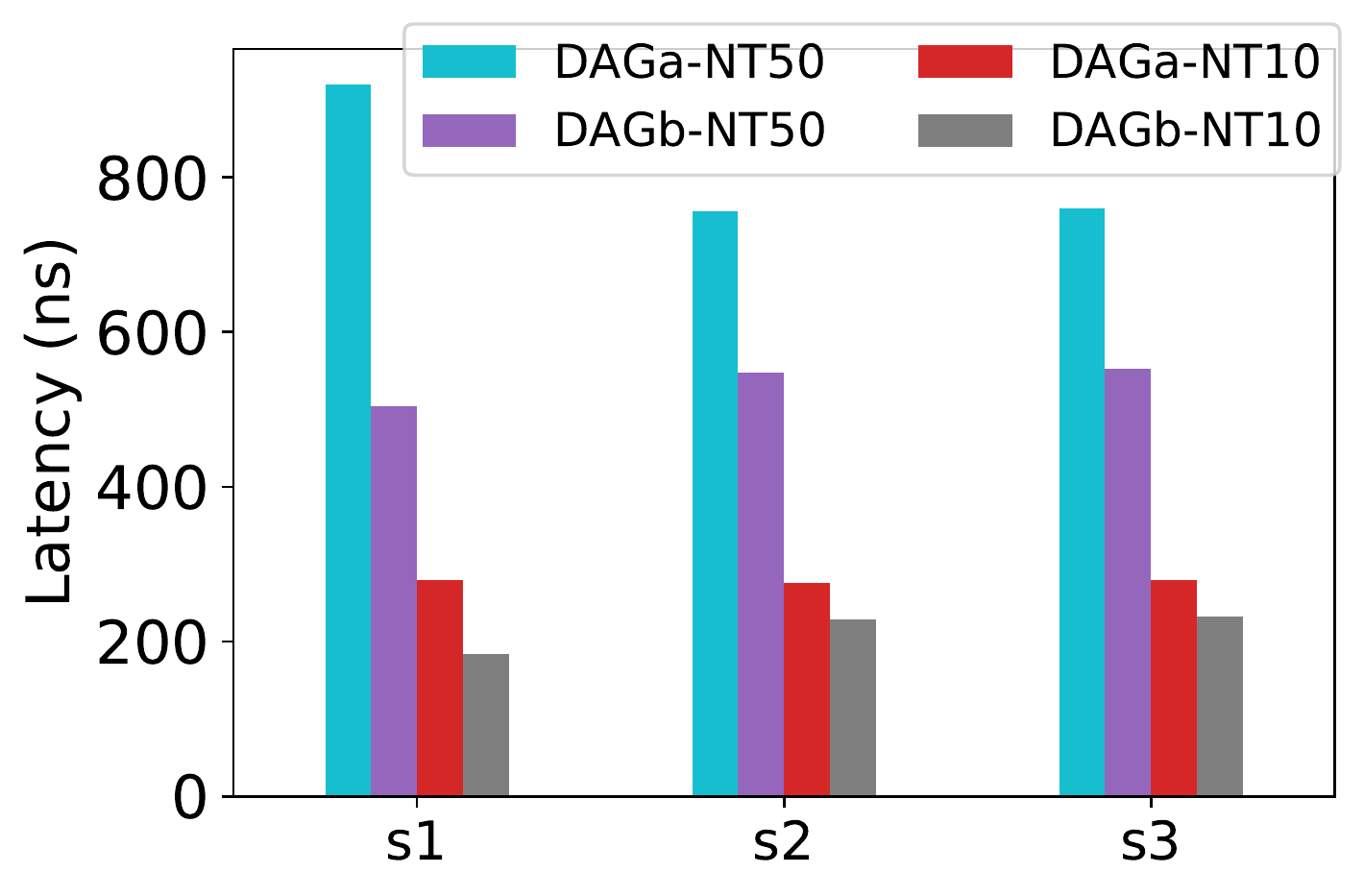}}
\vspace{-0.13in}
\mycaption{fig-dummy-nt-dag-lat}{\nt{} DAG Latency.}
{
S1-S3 same as Fig.10. NT50 and NT10 represent 50- and 10-cycle NTs.
}
\end{center}
\end{minipage}
\begin{minipage}{1.75in}
\begin{center}
\centerline{\includegraphics[width=\columnwidth]{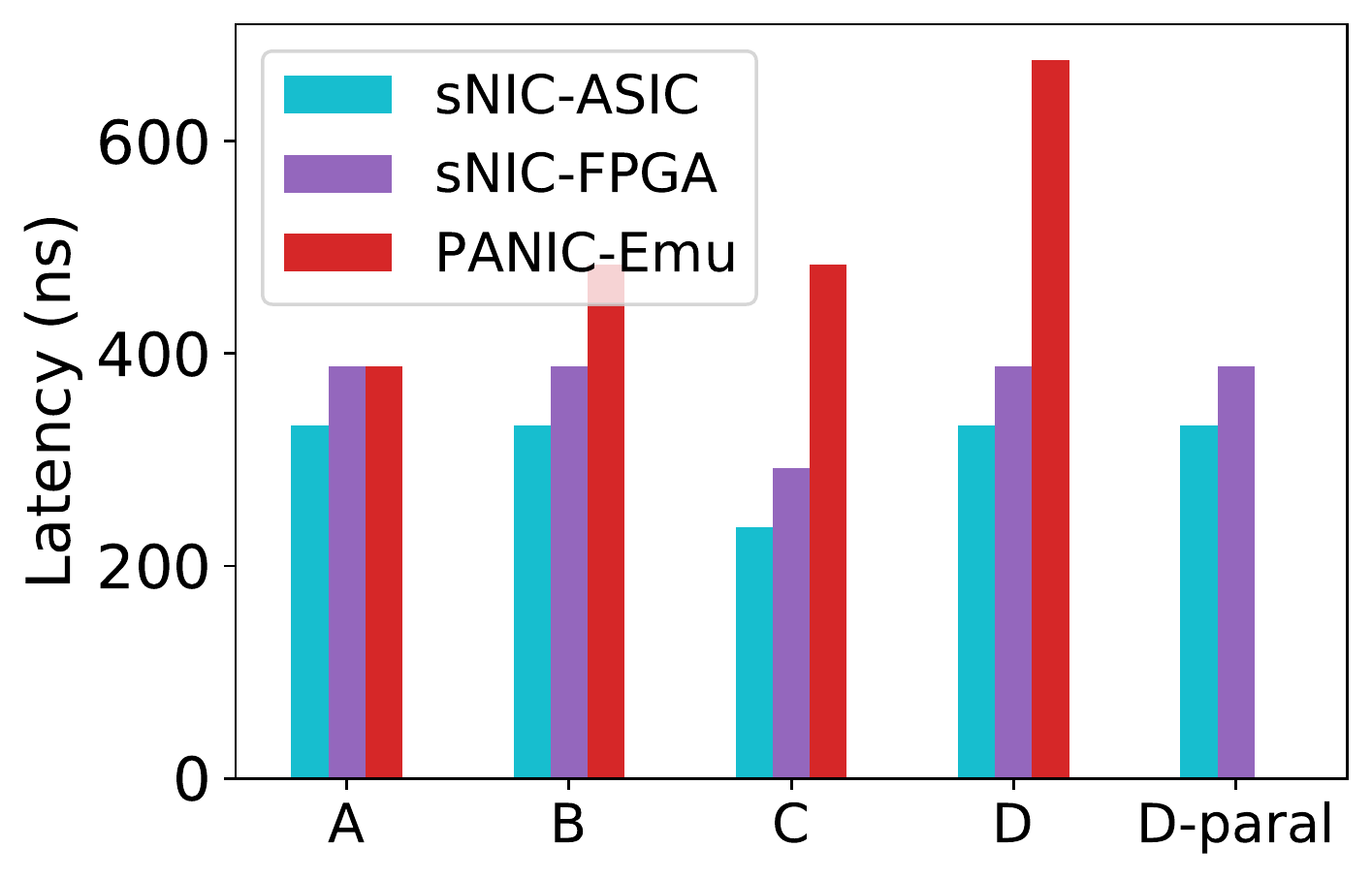}}
\vspace{-0.13in}
\mycaption{fig-nt-chain-lat}{Real \nt{} DAG Latency.}
{
\revise{sNIC-ASIC shows projected ASIC performance.}
}
\end{center}
\end{minipage}
\begin{minipage}{1.75in}
\begin{center}
\centerline{\includegraphics[width=\columnwidth]{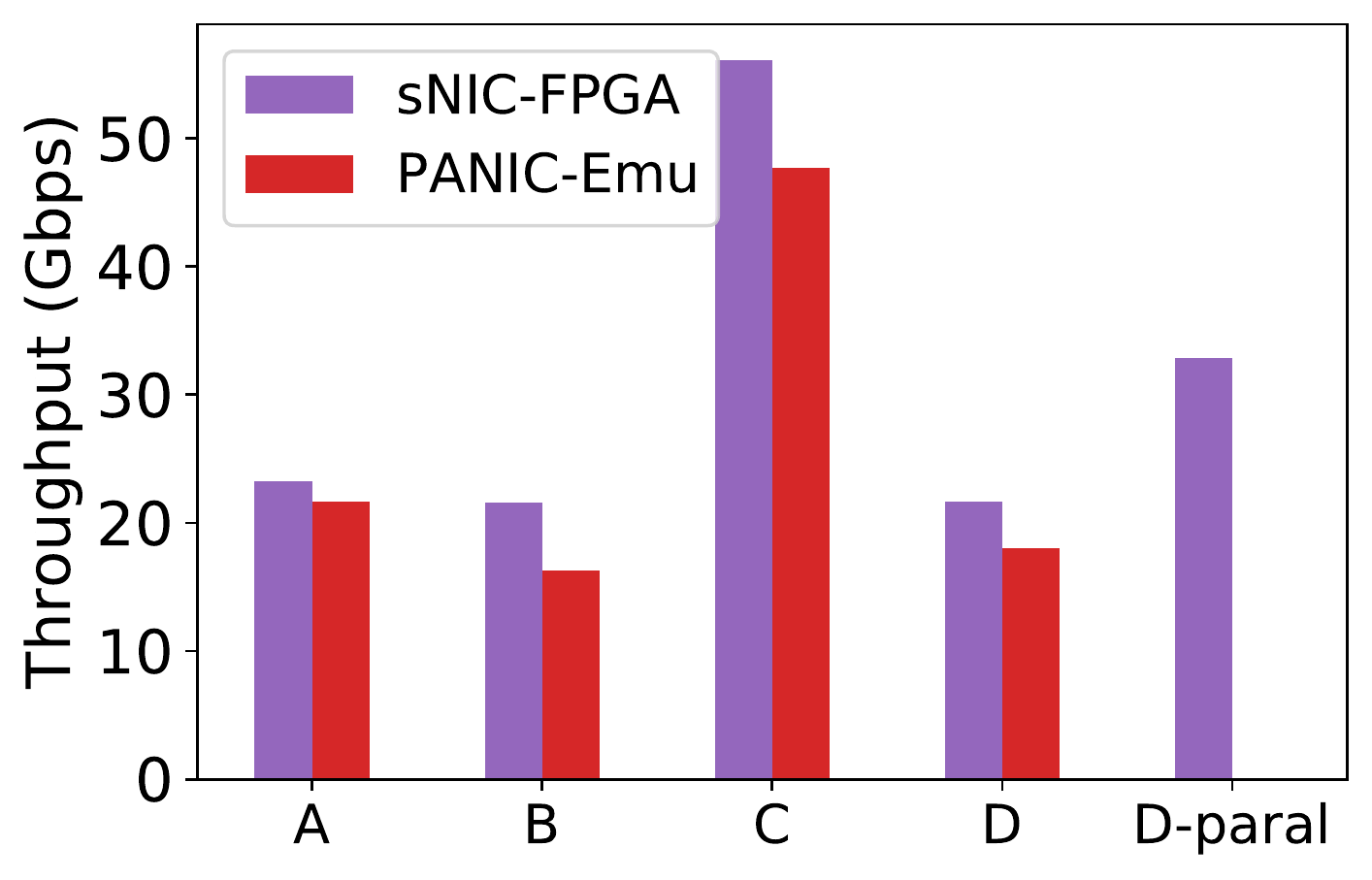}}
\vspace{-0.13in}
\mycaption{fig-nt-chain-tput}{Real \nt{} DAG Throughput.}
{
}
\end{center}
\end{minipage}
\vspace{-0.05in}
\end{figure*}
}

\bolditpara{\nt\ chaining.}~~
To evaluate the effect of \snic's \nt-chaining technique and compare it with PANIC, we use an artificial sequence of dummy \nt{}s with length from 2 to 7 (as prior work found real NT chains are usually less than 7 NTs~\cite{NFP-sigcomm17}).
We also evaluate a case where \snic\ splits the chain into two sub-chains.
Figure~\ref{fig-nt-chain} shows the total latency of running the \nt\ sequence with these schemes. 
\snic\ outperforms PANIC because it only goes through the scheduler once (for Single-Chain) or twice (for Half-Chain) for the entire chain, whilst Panic could go through the scheduler after every single NT in the chain.

\bolditpara{DAG parallelism and instance parallelism.}~~
We evaluate the different parallelism mechanisms introduced in \S\ref{sec:parallel} by measuring the throughput and latency of the three schemes (S1, S2, S3) in Figure~\ref{fig-pipeline} for the two DAGs: DAGa and DAGb.
Here, we treat all NTs as dummy ones that simply spins for 10 or 50 cycles for each packet; we set each NT's max processing bandwidth to 64\Gbps.
Figure~\ref{fig-dummy-nt-dag-tput} plots the throughput of packets going to the two DAGs under the three schemes. 
As can be seen, S3 improves the throughput of DAGa as we launch two instances of NT1, NT2, and NT4. The throughput is less than doubles of S1/S2's because the two instances of NT2 and NT4 are shared by DAGb.

Figure~\ref{fig-dummy-nt-dag-lat} plots the average execution time of the two DAGs under the three schemes and two different NT processing latency.
Adding DAG parallelism (S2) largely reduces the total execution time for DAGa when each NT runs for 50 cycles. However, when each NT runs for 10 cycles, S2 has no execution-time improvement. This is because S2 requires a packet to go back to the scheduler after processing NT3, which adds 16 cycles and is relatively large for short-running NTs. This indicates that when NTs are short running, it is more beneficial to chain them in a single region (\S\ref{sec:nt}).


We also evaluate DAG parallelism by increasing the number of dummy \nt{}s.
We compare two settings of \snic\ and our emulated PANIC.
The first setting runs all \nt{}s in parallel; the second setting splits \nt{}s into two groups and run these groups as two parallel \nt-chains (half-parallel).
Figure~\ref{fig-nt-parallel} shows the total latency of these schemes.
As expected, running all \nt{}s in parallel achieves the best performance.
Half-parallel only uses two regions and still outperforms PANIC. 

\bolditpara{Real \nt\ DAGs.}
We now present real \nt{} DAG evaluation results (DAGs in Figure~\ref{fig-nt-dag}). Figures~\ref{fig-nt-chain-lat} and \ref{fig-nt-chain-tput} plot the latency and throughput of running these DAGs on \snic\ and on our emulated PANIC. 
Similar to the artificial workload, real \nt{} DAGs also benefit from \snic, both in latency and in end-to-end burst throughput, especially when the DAG is long.
We also evaluate the impact of instance parallelism with a real \nt\ DAG (DAG-D). In this DAG, AES is the bottleneck \nt, which achieves the lowest bandwidth. Thus, by launching two parallel AES \nt{}s in DAG-D-paral, we see a clear throughput increase in Figure~\ref{fig-nt-chain-tput}.

\bolditpara{\nt\ sharing.}~~
We evaluate the effect of sharing \nt{}s by running DAG-D in Figure~\ref{fig-nt-dag} as a background workload, with bandwidth consumption of 15\Gbps.
We then add one of the DAG-A/C as the foreground workload and vary the foreground workload's traffic load.
We compare the foreground workload's throughput and FPGA area consumption when enabling and disabling \nt\ sharing.
We calculate FPGA area consumption by measuring FPGA LUTs and BRAM each \nt\ consumes, normalized to DAG-D-paral's usage.
Figure~\ref{fig-nt-sharing-tput} plots the throughput of DAG-A and DAG-C divided by their area consumed. 
Overall, sharing improves throughput per area for both DAGs, since without sharing, we would need to launch each DAG in its own region.
DAG-A's throughput keeps increasing as its load increases until 15\Gbps. Afterwards, its throughput saturates at 15\Gbps. This is because the AES NT can only achieve 30\Gbps\ maximum throughput, and the background DAG-D consumes 15\Gbps. Yet, we still see a small gain in throughput per area. DAG-C does not utilize AES and can fully utilize the unused bandwidth of the remaining NTs in DAG-D. Thus, it can scale to 60\Gbps, providing a large benefit in throughput per area with sharing.

\bolditpara{Effect of victim cache.}~~
To evaluate the effect of our victim-cache design (\S\ref{sec:nt-deploy}), we set the baseline to be disabling victim cache (blue dot at the middle of Figure~\ref{fig-sim-single-snic}).
We then change how often a de-scheduled \nt\ can be kept around as a victim instead of being completely deleted (shown as percentage on the green line). 
This models how often an \snic's area is free to host victim \nt{}s.
As expected, the more de-scheduled \nt{}s we keep around, the better performance we achieve, with no victim cache (baseline) having the worst performance.
The OpEx implication is less intuitive.
Here, we only count the time and amount of \nt\ regions that are actually accessed, as only those will cause the dynamic power (when idle, FPGA has a static power consumption regardless of how it is programmed).
With fewer de-scheduled \nt{}s kept around, more \nt{}s need to be re-launched (through FPGA PR) when the workload demands them. 
These re-launching overhead causes the OpEx to also go up.

\bolditpara{Effect of area over-commitment.}~~
We change the degree of area over-commitment by limiting how much hardware resources (\ie, NT regions) the workload can use compared to the ideal amount of resources needed to fully execute it.
Figure~\ref{fig-sim-single-snic} shows that increasing the area over-commitment rate causes worse performance but less resources used. 


{
\begin{figure*}[th]
\begin{minipage}{0.7\columnwidth}
\begin{center}
\centerline{\includegraphics[width=\columnwidth]{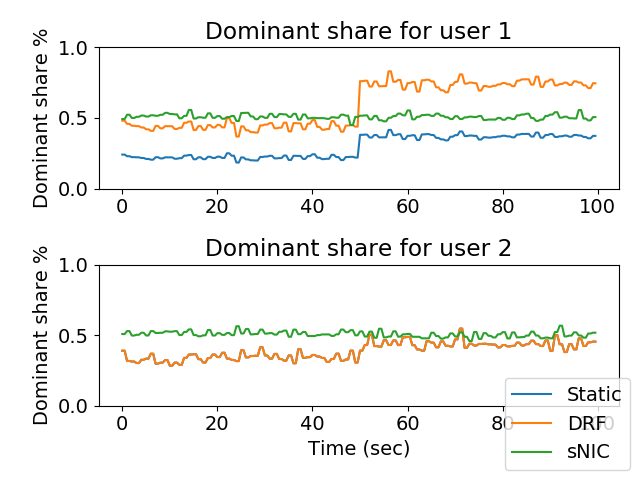}}
\vspace{-0.1in}
\mycaption{fig-dom}{Dominant Resource Share.}
{The Static line overlaps with DRF for user 2.}
\end{center}
\end{minipage}
\begin{minipage}{0.8\columnwidth}
\begin{center}
\centerline{\includegraphics[width=0.9\columnwidth]{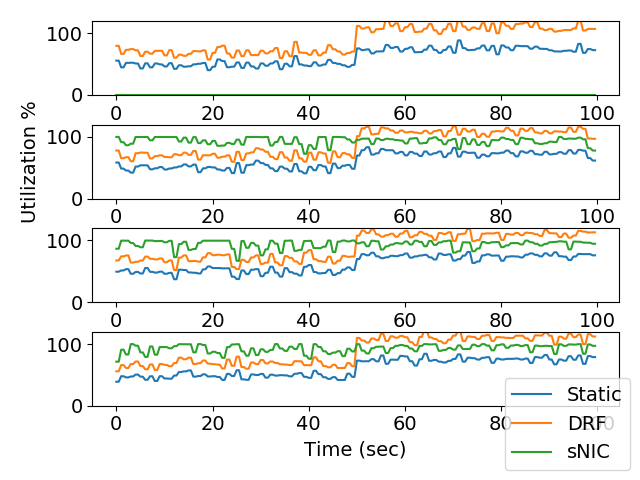}}
\vspace{-0.15in}
\mycaption{fig-util}{Resource Utilization.}
{\small{Starting from the top, each row shows the utilization of FPGA fabric, ingress bandwidth, egress bandwidth, and memory space. The first row's \snic\ line overlaps with DRF's.}}
\end{center}
\end{minipage}
\begin{minipage}{0.55\columnwidth}
\begin{center}
\centerline{\includegraphics[width=\columnwidth]{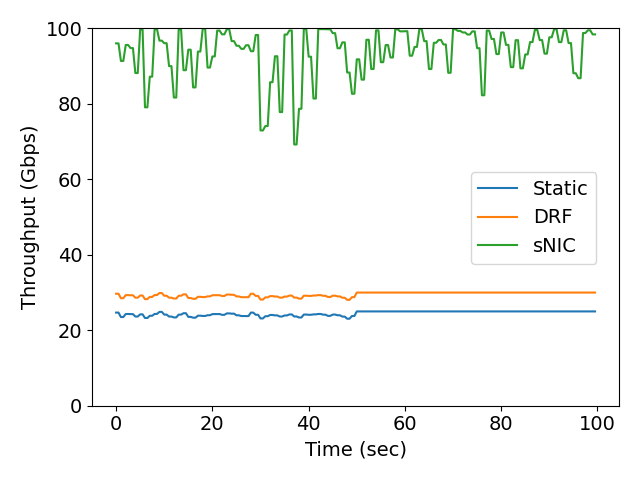}}
\vspace{-0.1in}
\mycaption{fig-tput}{Aggregated Throughput}{}
\end{center}
\end{minipage}
\vspace{-0.15in}
\end{figure*}
}

\bolditpara{Fair resource sharing.}
To evaluate the effectiveness of our fairness policy, 
we ran a synthetic workloads that includes two users and in a multi-resource environment. User 1 runs 4 dummy \nt{}s in a chain, and user 2 runs 2 dummy \nt{}s in a chain. User 2's chain is a subset of user 1's. Their user-supplied load requirement is the same. Thus, a good fairness policy should ensure that they each gets half of their dominant resource. We run the two workloads for 100 seconds. At 50 seconds, user 1's load increases.
We evaluate this workload on three different schemes. \emph{Static} is the baseline, where each user gets assigned an equal number of \nt\ regions.
The \emph{DRF} scheme uses DRF to space share, but does not allow the time-sharing of \nt\ DAGs amongst different users.
Finally, \emph{\snic} is our complete \snic\ fairness policy.

{
\begin{figure*}[th]
\begin{minipage}{1.75in}
\begin{center}
\centerline{\includegraphics[width=\columnwidth]{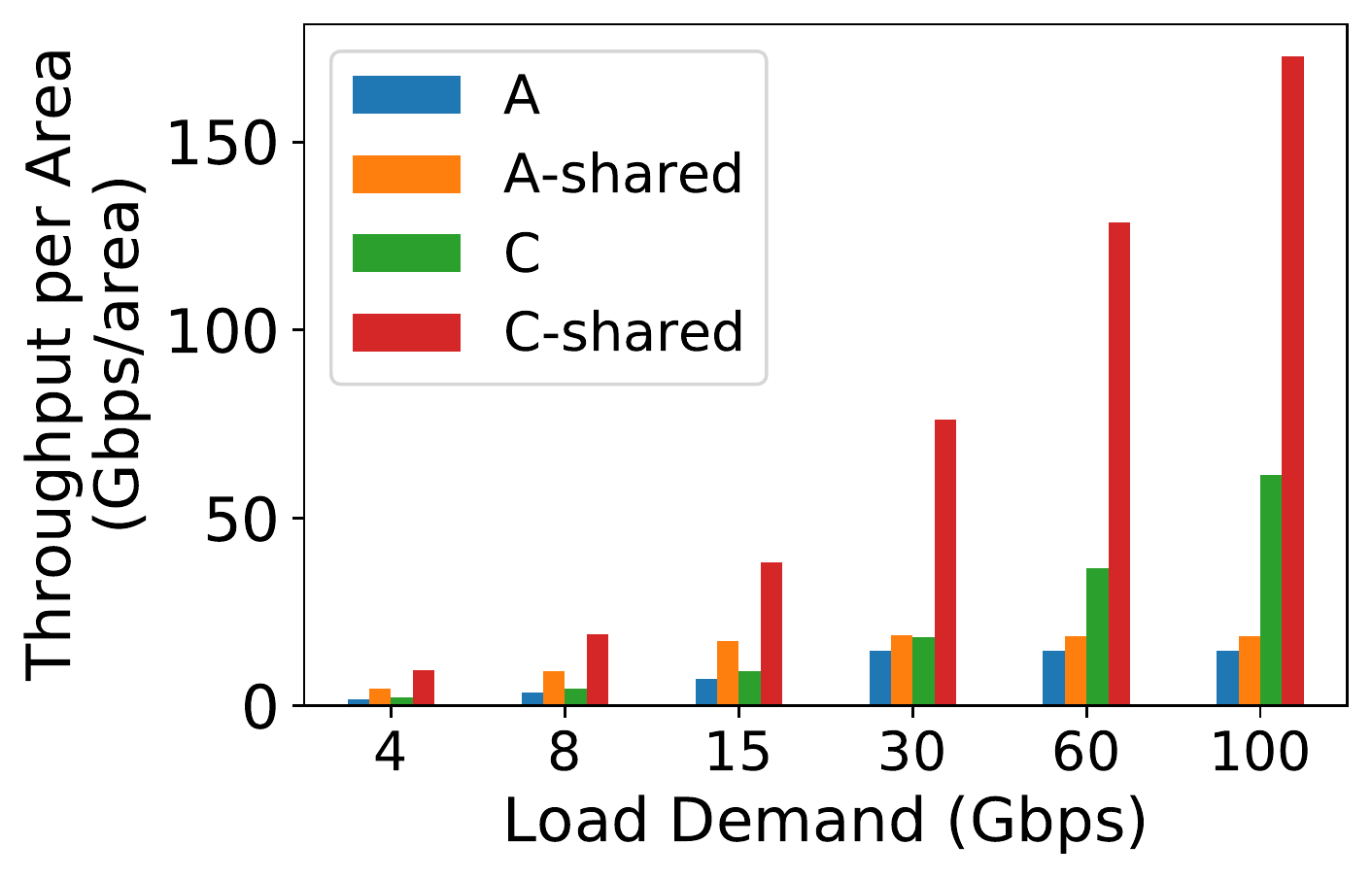}}
\vspace{-0.15in}
\mycaption{fig-nt-sharing-tput}{NT sharing}
{
A and C sharing chain D which is running at 15gbps.
}
\end{center}
\end{minipage}
\begin{minipage}{1.75in}
\begin{center}
\centerline{\includegraphics[width=\columnwidth]{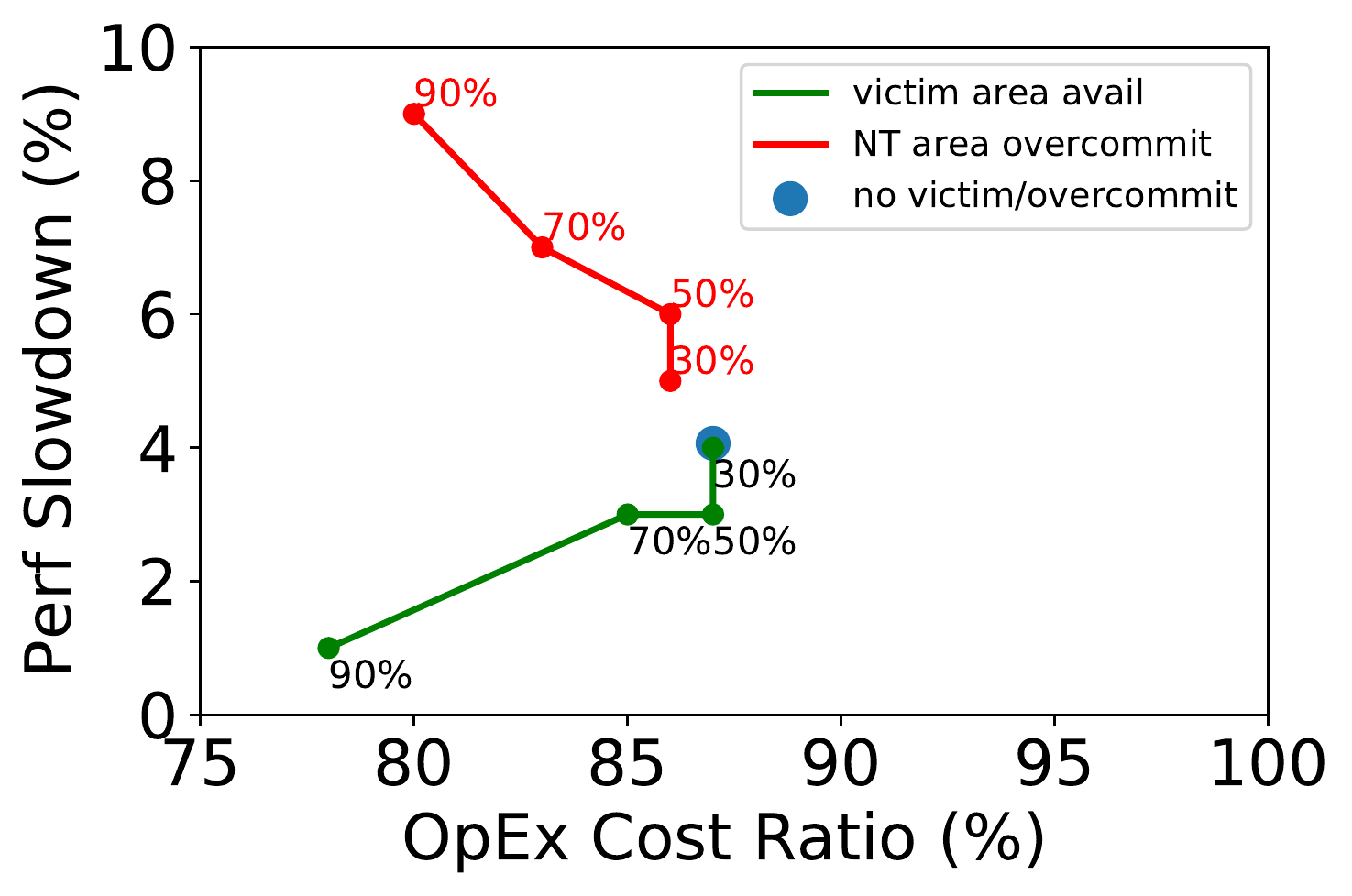}}
\vspace{-0.15in}
\mycaption{fig-sim-single-snic}{Single \snic{} Sensitivity.}
{
Each lines is running a different set of experiments.
}
\end{center}
\end{minipage}
\begin{minipage}{1.75in}
\begin{center}
\centerline{\includegraphics[width=\columnwidth]{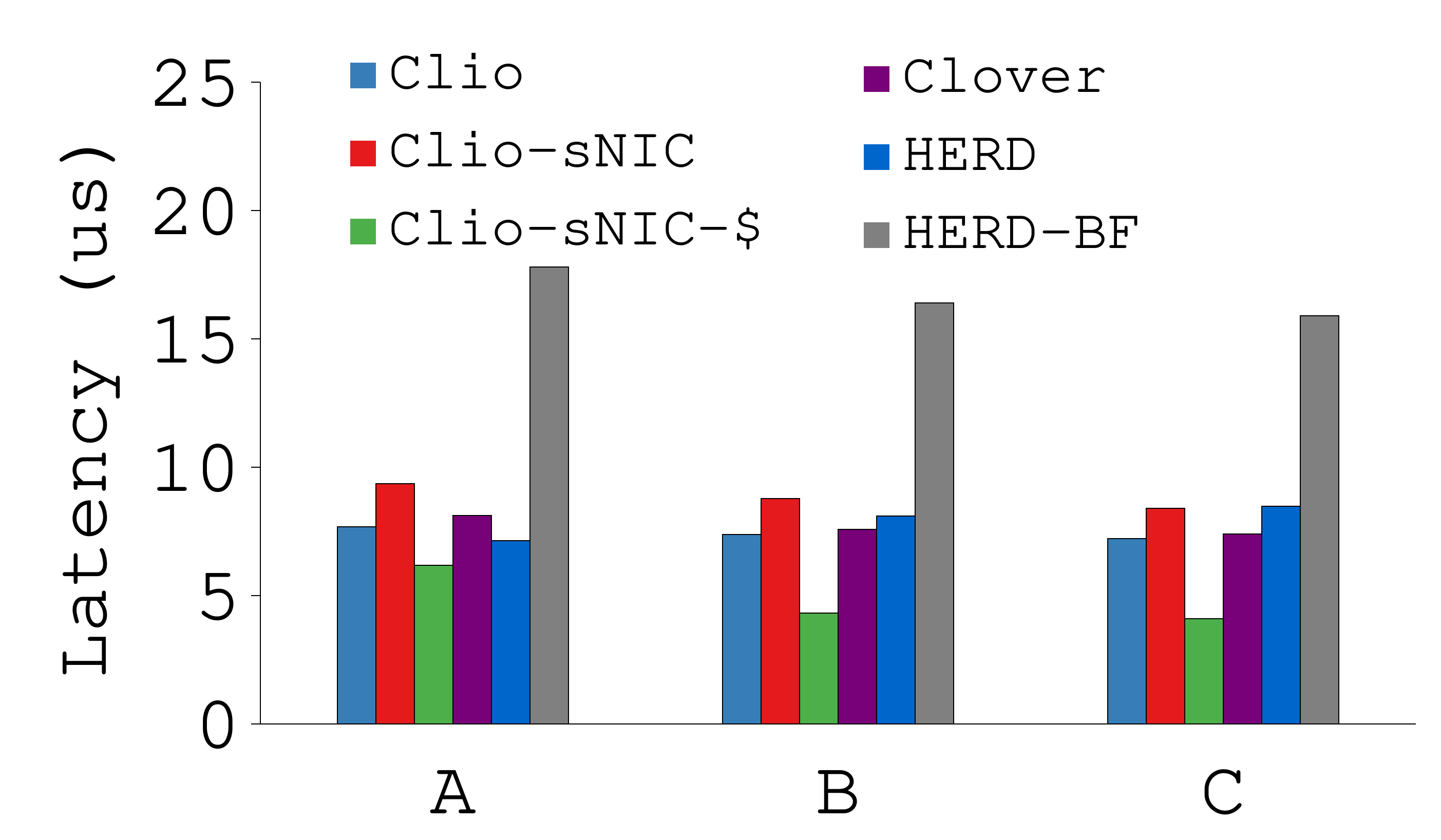}}
\vspace{-0.1in}
\mycaption{fig-ycsb}{\small YCSB Latency.}
{
}
\end{center}
\end{minipage}
\begin{minipage}{1.75in}
\begin{center}
\centerline{\includegraphics[width=\columnwidth]{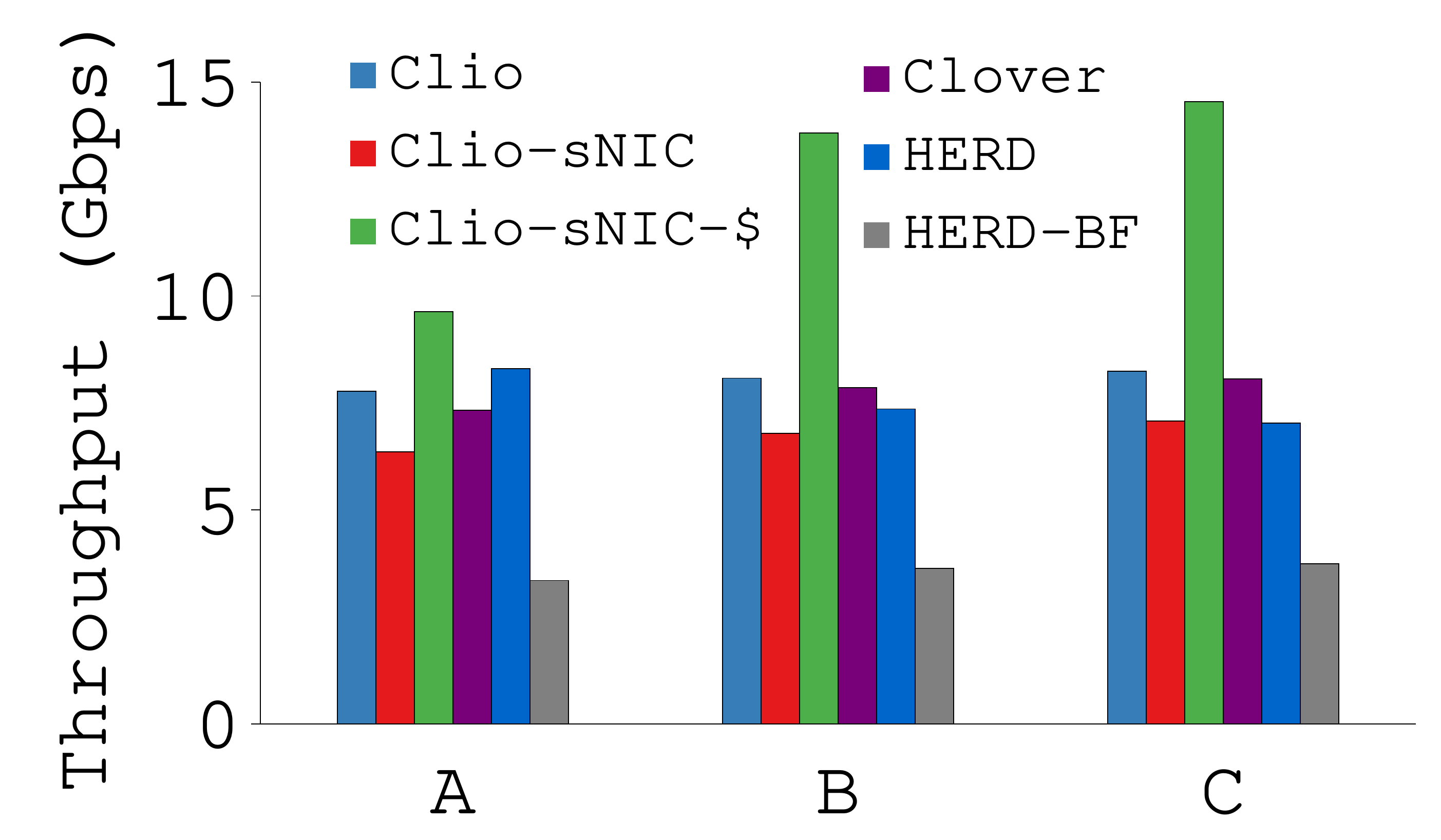}}
\vspace{-0.1in}
\mycaption{fig-ycsb-tput}{\small YCSB Throughput.}
{
}
\end{center}
\end{minipage}
\vspace{-0.1in}
\end{figure*}
}

Figure~\ref{fig-dom} shows the resulting dominant share timeline for the two users, and Figure~\ref{fig-util} shows the  utilization of each type of resource aggregated across the two users. 
\emph{\snic} consistently delivers fair share for both users even when one user's load changes.
In contrast, the \emph{DRF} scheme cannot adjust to the load change, because it statically decides resource allocation.

Figure~\ref{fig-tput} shows the aggregated throughput of the two users.
By allowing the time sharing of common \nt\ DAGs, \emph{\snic} allows for underutilized DAGs to process other users' flows. Thus, \emph{\snic} achieves higher aggregated throughput than DRF. Compared to \emph{Static}, \emph{DRF} can fully use all the regions, resulting in a slightly higher aggregated throughput \emph{Static}.


\subsection{End-to-End Application Performance}

We now present our end-to-end application performance. 
For this experiment, we deployed \snic\ with Clio~\cite{Clio}, a recent disaggregated memory system.
Clio includes a client-side server that issues remote-memory access requests such as key-value get and put. It also includes a network-attached hardware-based memory device that hosts the user data and performs the accesses.
When deploying \snic, we connect the \snic\ board to the Clio memory board via Ethernet. We then connect the client-side server and the \snic\ board to a 100\Gbps\ Ethernet switch.
After the connection, we offload Clio's transport (go-back-N) to \snic. We further deploy a key-value cache \nt\ in the \snic.

We run YCSB's workloads A (50\% set, 50\% get), B (5\% set, 95\% get), and C (100\% get)~\cite{ycsb-socc10} for this experiment. We use 100K key-value entries and run 100K operations per test, with YCSB's default key-value size of 1\KB\ and Zipf accesses ($\theta=0.99$). 
We compare \snic\ to several baselines: the original Clio, a one-sided RDMA-based key-value store, Clover~\cite{Tsai20-ATC}, a RPC-like RDMA-based key-value store, HERD~\cite{Kalia14-RDMAKV}. We run Clover and HERD with NVidia 100\Gbps\ ConnectX-4 RDMA NIC. We also run HERD on the NVidia 100\Gbps\ BlueField-Gen1 SmartNIC.

We evaluate the \snic\ performance when we only run the Go-back-N transport \nt\ at it and when we run both the transport \nt\ and the key-value caching \nt.
Figures~\ref{fig-ycsb} and ~\ref{fig-ycsb-tput} show the latency and throughput of \snic\ and other baseline systems.
\snic's performance is on par with Clio, Clover, and HERD, as it only adds a small overhead to the baseline Clio.
With caching \nt, \snic\ achieves the best performance among all systems, esp. on throughput. 
This is because all links in our testbed are 100\Gbps\ except for Clio board's 10\Gbps\ link, which connects to the \snic\ board. When there is a cache hit at the \snic, we avoid going to the 10\Gbps\ Clio boards.
HERD-BF performs the worst because of the slow link between its NIC and the ARM processor. Newer generations of BlueField are more powerful than BlueField-1. Unfortunately, we do not have access to the newer generations. 

\section{Conclusion}
\label{sec:conclude}

We presented a multi-tenant, programmable, hardware-based SmartNIC, \snic, that focuses on optimizing \nt\ DAG execution.
Our proposal include the design of \nt\ chain, DAG and instance parallelism, various \nt\ launching optimization techniques, and a full set of fair sharing mechanisms.
Our FPGA prototype demonstrates the performance, fairness, and cost benefits of \snic.

\newpage

\clearpage
\begin{small}
\bibliographystyle{IEEEtranS}
\bibliography{all-defs,all,personal,all-confs,local,paper}
\end{small}

\clearpage

\appendix

\section{Appendix}

\subsection{FPGA Resource Utilization}

The following table shows the FPGA resources used by \snic{} shell.
Most of the resources are left for running \nt{}s.

\begin{center}
\scriptsize
\begin{tabular}{ p{0.6in} | p{0.2in} |p{0.27in} }
 & \textbf{Logic} & \textbf{Memory} \\
\textbf{Module} & \textbf{(LUT)} & \textbf{(BRAM)} \\
\hline
\hline
\snic{} Core & 4.36\%   & 4.74\% \\
Packet Store & 0.91\%   & 9.17\% \\
PHY+MAC      & 0.72\%   & 0.35\% \\
DDR4Controller         & 1.57\%   & 0.29\% \\
MicroBlaze   & 0.25\%   & 1.81\% \\
Misc         & 1.52\%   & 0.75\% \\
\hline
\textbf{Total}        & \textbf{9.33\%}   & \textbf{17.11\%} \\
\end{tabular}
\end{center}

\subsection{Cost Calculation}
We explain the different deployment models and the cost calculation formulas behind our CapEx comparisons.
We limit our scope to rack-scale as the higher-level network hierarchies
are orthogonal to the resource pool deployment models.
We calculate that, to deploy a certain number of endpoints, what's the
network cost (i.e., the network interface card, cable, and switch port costs).

We compare the following models:
1) Non-disaggregation model, or the traditional model, termed \texttt{traditional}.
2) Disaggregation model, in which we insert the network pool between endpoints and the ToR switch (Figure~\ref{fig-topology} (a)), termed \texttt{ring}.
3) Disaggregattion model, in which we connect the pool of network devices directly to the ToR switch (Figure~\ref{fig-topology} (b)), termed \texttt{direct}.
For both disaggregation models, we further compare two type of devices: sNIC which has auto-scaling capability and multi-host NIC which can only provision for max resource usage. With runtime dynamic scaling and load balancing features, sNICs can provision for less than the max required resource , the specific ratio is calculated by comparing a particular workload's the sum-of-peak versus the peak-of-sum.

In all, we have the following models under comparison:
\texttt{traditionl, sNIC-direct, sNIC-ring, mhnic-direct, mhnic-ring}.

We now detail the cost calculations.
In the traditional non-disaggregation model,
each endpoint has a full-fledged NIC and a normal high-speed cable for connection to the ToR switch.
In both disaggregation models, since most network tasks are offloaded to the network resource pool, each endpoint can uses a down-scaled NIC.
Furthermore, the last hop link layer between endpoints and the network resource pool is reliable, we can leverage down-scaled, cheaper and less reliable physical cable~\cite{RAIL-NSDI}.

We use the following parameters in our calculation:
\begin{itemize}
\item Deploy \texttt{N} devices.
\item Each switch port has a cost of \texttt{costSwitchPort}
\item A full-fledged NIC's cost is \texttt{costNIC}. A down-scaled NIC cost is \texttt{costDSNIC}.
\item A normal high-speed cable cost is \texttt{costCable}.
A down-scaled less reliable physical cable cost is \texttt{costDSCable}.
\item A consolidation ratio \texttt{consolidRatio} determines how many endpoints are sharing one network resource pool device. We can calculate the number of network pool devices by \texttt{M = N / consolidRatio}.
\item For a network device, only a certain portion is dedicated to running network task, other parts are used as shell. We define the cost ratio used by network task to be \texttt{NTCostRatio}.
\item The peak-of-sum versus the sum-of-peak yields the auto-scaling potentials. A multi-host NIC (mhnic) provisions for the sum-of-peak while an sNIC provisions for the peak-of-sum. We call this ratio \texttt{capExConsolidRatio}.
\item The multi-host NIC's cost can be calculated as \texttt{costMHNIC = costNIC * N}.
\item The sNIC's cost can be calculated as \texttt{costsNIC = costMHNIC * capExRatio}, in which \texttt{capExRatio = (1 - NTCostRatio) + NTCostRatio * capExConsolidRatio}.
\end{itemize}

We now define each model's cost.

The traditional deployment model's cost is straightforward, it includes NIC, cable and switch ports:
\begin{gather}
N * (costNIC + costCable + costSwitchPort)
\end{gather}

The disaggregation models' cost has more moving parts than the traditional. It includes the down-scaled NICs and cables, network pool devices, the cables to the ToR switch, and switch ports.

The first disaggregation model (Figure~\ref{fig-topology} (a)) can be calculated as follows (for both \texttt{sNIC-ring, mhnic-ring}). 
\begin{align}
N * (costDSNIC + costDSCable) + \\
M * (costsNIC + costCable + costSwitchPort)
\end{align}

The second disaggregation model (Figure~\ref{fig-topology} (b)) can be calculated as follows (for both \texttt{sNIC-direct, mhnic-direct}).
\begin{align}
N * (costDSNIC + costCable + costSwitchPort) + \\
M * (costsNIC + costCable + costSwitchPort)
\end{align}

This tables shows the real-world numbers we use.

\begin{center}
\scriptsize
\begin{tabular}{|l|l|l|} 
 \hline
 Parameters & Value & Note \\
 \hline\hline
 costSwitchPort & \$250 & FS 100Gbps switch~\cite{fs-64port-switch} \\
 costNIC & \$500 & Mellanox Connect-X5 \\
 costCable & \$100 & FS DAC 100Gbps cable \\
 costDSNIC & costNIC * 0.2 & Numbers from our prototpe \\
 costDSCable & costCable * 0.6 & ~\cite{RAIL-NSDI} \\
 consolidRatio & 4 & Current model\\
 NTCostRatio & 0.9 & Numbers from our prototype \\
 capExConslidRatio & 0.23 & Facebook Hadoop trace~\cite{facebook-sigcomm15} \\
 \hline
\end{tabular}
\end{center}


{
\begin{figure*}[th]
\begin{minipage}{\figWidthSix}
\begin{center}
\centerline{\includegraphics[width=\columnwidth]{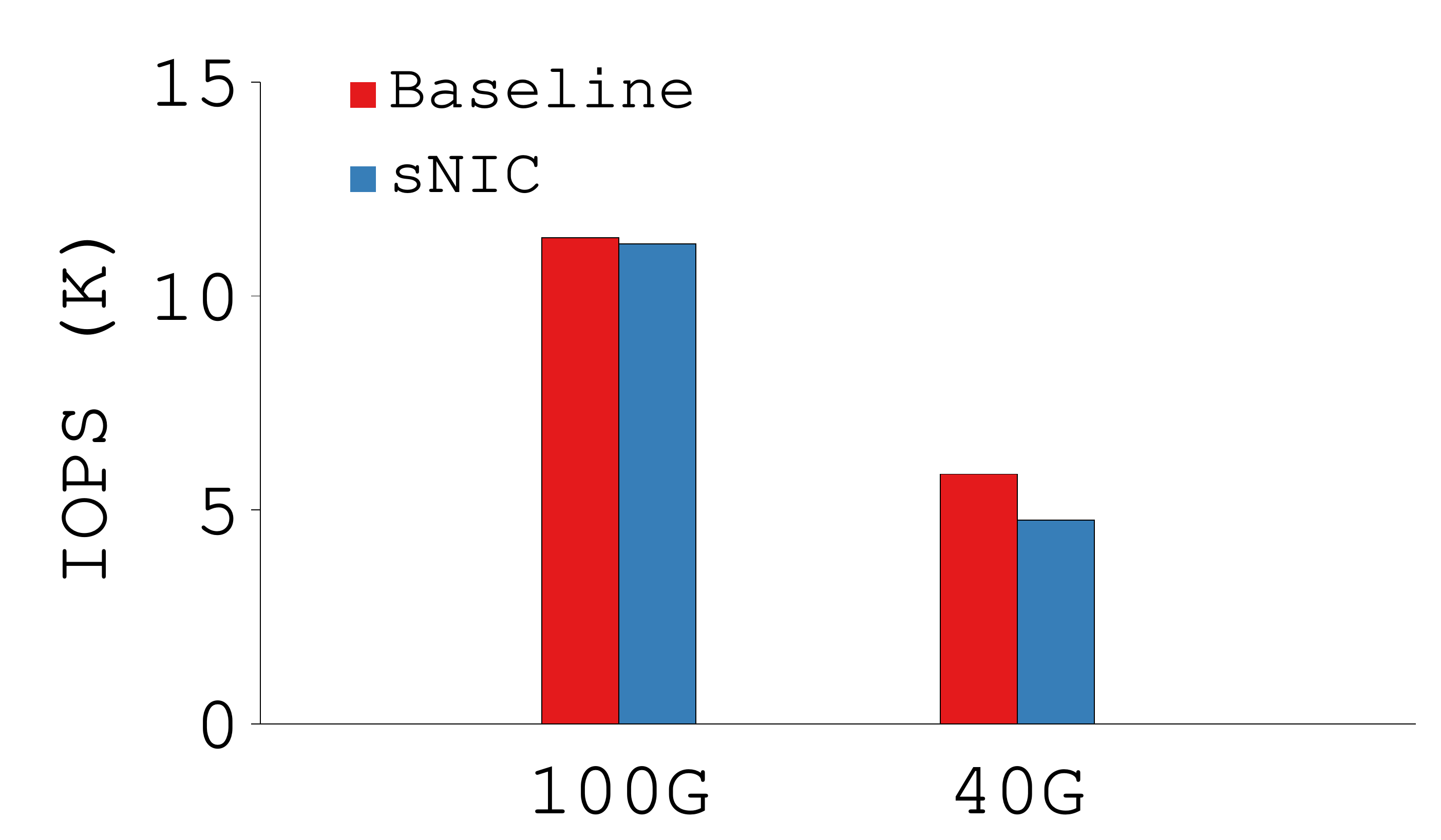}}
\vspace{-0.1in}
\mycaption{fig-kv-consolid}{Consolidation Performance w/ FB Key-Value.}
{
}
\end{center}
\end{minipage}
\begin{minipage}{\figWidthSix}
\begin{center}
\centerline{\includegraphics[width=\columnwidth]{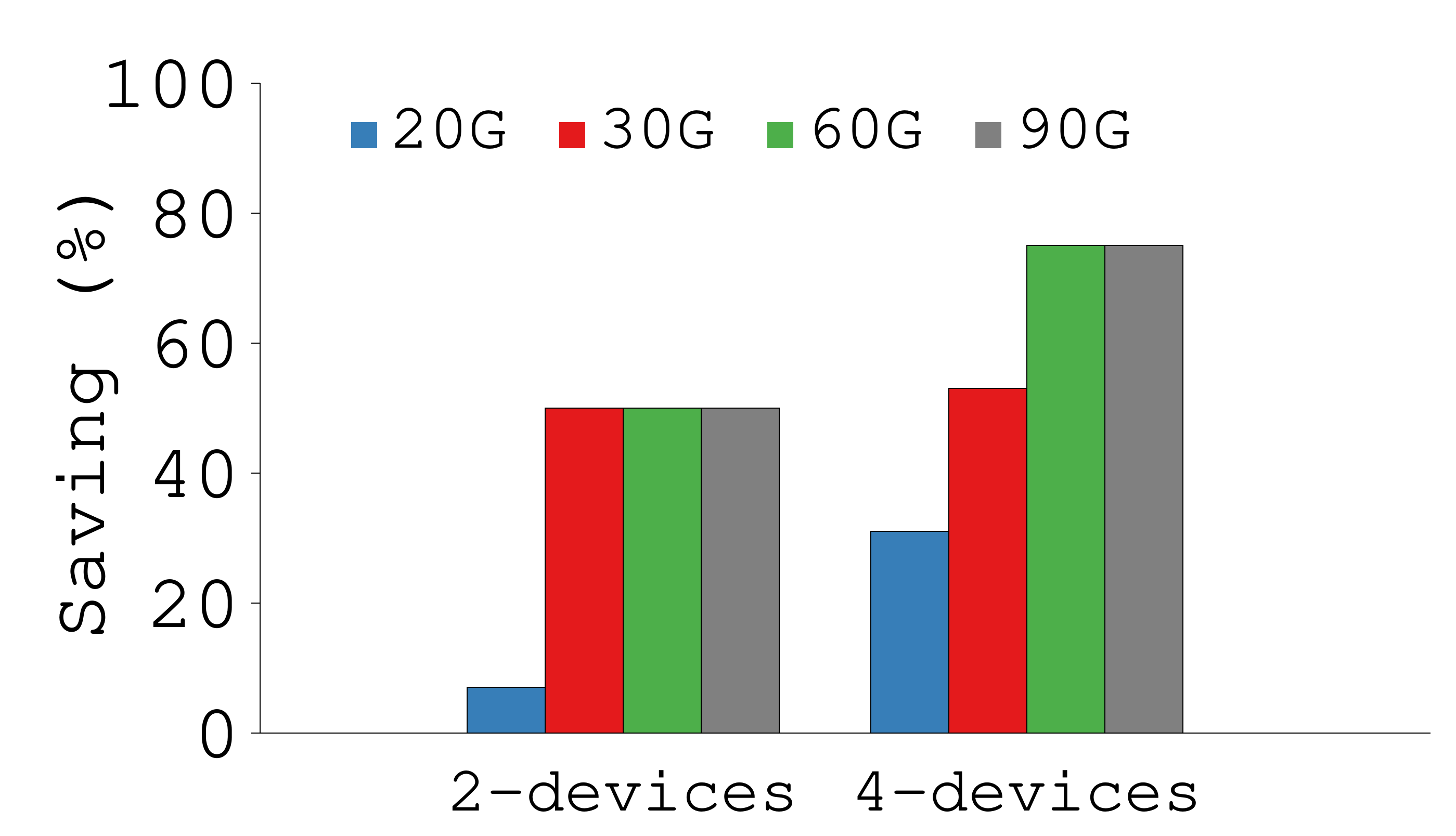}}
\vspace{-0.1in}
\mycaption{fig-kv-cost}{Consolidation Resource Usage w/ FB KV.}
{
}
\end{center}
\end{minipage}
\vspace{-0.1in}
\end{figure*}
}
\subsection{End-to-End Application Performance and Cost with Consolidation}

To evaluate the benefit and tradeoff of consolidation, we deploy a testbed with four sender and four receiving servers with four setups:
each endhost connects to a ToR switch with 100\Gbps\ or 40\Gbps\ link (baseline, no consolidation), and four endhosts connect to an \snic, each with 100\Gbps\ or 40\Gbps\ link, and the \snic\ connects to the ToR switch with a 100\Gbps\ or 40\Gbps\ link (\snic\ consolidation).
For both settings, we execute two \nt{}s, firewall and NAT, in FPGA. 
For the baseline, each endhost has its own set of \nt{}s, while 
\snic\ autoscales \nt{}s as described in \S\ref{sec:policy}.
On each server, we generate traffic to follow inter-arrival and size distribution reported in the Facebook 2012 key-value store trace~\cite{Atikoglu12-SIGMETRICS}.

Figure~\ref{fig-kv-consolid} reports the throughput comparison of \snic\ and the baseline.
\snic\ only adds 1.3\% performance overhead to the baseline under 100\Gbps\ network and 18\% overhead under 40\Gbps\ network. 
We further analyze the workload and found its median and 95-percentile loads to be 24\Gbps\ and 32\Gbps.
With four senders/receivers, the aggregated load is mostly under 100\Gbps\ but often exceeds 40\Gbps.
Note that a multi-host NIC would not be able to achieve \snic's performance, as it subdivides the 100\Gbps\ or 40\Gbps\ into four 25\Gbps\ or 10\Gbps\ sub-links, which would result in each endhost exceeding its sub-link capacity.

We then calculate the amount of FPGA used for running the \nt{}s multiplied by the duration they are used for, to capture the run-time resource consumption with \snic's autoscaling mechanism. The baseline has one set of \nt{}s per endhost for the whole duration.
Figure~\ref{fig-kv-cost} shows this comparison when consolidating two and four endhosts to an \snic\ and using \nt{}s of different performance metrics.
For a slower \nt{} (\eg, one that can only sustain 20\Gbps\ max load), the \snic\ auto-scales more instances of it, resulting in less cost saving.
Our implementation of firewall \nt{} reaches 100\Gbps, while the AES \nt\ is 30\Gbps, resulting in a 64\% cost saving when deploying both of them.

\end{document}